\documentclass[a4paper]{article}
\pdfoutput=1
\usepackage{a4wide}
\usepackage{booktabs}
\usepackage{graphicx,color,epstopdf}
\usepackage{amsmath,amssymb,amsfonts}
\usepackage{hyperref}
\usepackage{array}
\usepackage[small,bf]{caption}
\usepackage{color}
\usepackage[usenames,dvipsnames]{xcolor}
\usepackage{cite}
\usepackage{bbm}

\newcommand{\gev}{\ensuremath{\,\mathrm{GeV}}}
\newcommand{\tev}{\ensuremath{\,\mathrm{TeV}}}

\newcommand{\eqn}[1]{&\hspace{-0.6em}#1\hspace{-0.6em}&}

\definecolor{DeepPink}{rgb}{1.0, 0.08, 0.58}

\allowdisplaybreaks

\begin{document}

\begin{titlepage}

\vspace*{-15mm}
\begin{flushright}
NCTS-PH/1722
\end{flushright}
\vspace*{0.7cm}

\begin{center}

{ \bf\LARGE A supersymmetric electroweak scale \\[2mm] seesaw model } 
\\[8mm]
Jung Chang$^{\, a,b}$ \footnote{E-mail: \texttt{lovejesus99wwjd@gmail.com}},
Kingman Cheung$^{\, b,c,d,}$ \footnote{E-mail: \texttt{cheung@phys.nthu.edu.tw}},
Hiroyuki Ishida$^{\, b,}$ \footnote{E-mail: \texttt{hiroyuki403@cts.nthu.edu.tw}},
Chih-Ting Lu$^{\, c,}$ \footnote{E-mail: \texttt{timluyu@hotmail.com}}, \\
Martin Spinrath$^{\, b}$ \footnote{E-mail: \texttt{martin.spinrath@cts.nthu.edu.tw}}
and Yue-Lin Sming Tsai$^{\, b}$ \footnote{E-mail: \texttt{sming.tsai@cts.nthu.edu.tw}}
\\[1mm]
\end{center}
\vspace*{0.50cm}
\centerline{$^{a}$ \it Department of Physics, Chonnam National University,
300 Yongbong-dong,}
\centerline{\it Buk-gu, Gwangju, 500-757, Republic of Korea}
\vspace*{0.2cm}
\centerline{$^{b}$ \it Physics Division, National Center for Theoretical Sciences, Hsinchu 30013, Taiwan}
\vspace*{0.2cm}
\centerline{$^{c}$ \it Department of Physics, National Tsing Hua University, 
Hsinchu 30013, Taiwan}
\vspace*{0.2cm}
\centerline{$^{d}$ \it Division of Quantum Phases \& Devices, School of Physics, 
Konkuk University,}
\centerline{\it Seoul 143-701, Republic of Korea}
\vspace*{1.20cm}

\begin{abstract}
\noindent
In this paper we propose a novel supersymmetric inverse
seesaw model which has only one additional $Z_6$ symmetry.
The field content is minimal to get a viable neutrino spectrum
at tree-level.
Interestingly, the inverse seesaw scale in our model is related to the scale
of electroweak symmetry breaking. 
Due to that origin we are less biased about hierarchies
and discuss three different types of the inverse seesaw mechanism
with different phenomenologies.
We can successfully reproduce neutrino masses and mixing and
our model is consistent with current bounds on neutrinoless double
beta decay, non-unitarity of the PMNS matrix and charged lepton
flavor violation.
\end{abstract}

\end{titlepage}
\setcounter{footnote}{0}

\renewcommand{\thefootnote}{\arabic{footnote}}

\section{Introduction}
\label{sec:introduction}

Supersymmetry (SUSY) is still one of the most attractive models for
physics beyond the Standard Model (SM). It not only solves the gauge hierarchy
problem, but also provides a dynamical mechanism for electroweak
symmetry breaking.
The minimal supersymmetric extension of the SM, called
the minimal supersymmetric standard model (MSSM) \cite{mssm}, has 
roughly doubled the degrees of freedom of the SM and has many phenomenological 
implications for, e.g.\ Higgs and flavor physics.
The MSSM is usually defined with an ad-hoc $Z_2$ symmetry, known
as $R$-parity, which can provide
a dark matter (DM) candidate to explain the DM relic density of the
universe.

Another undeniable evidence of physics beyond the SM is neutrino
masses and oscillations \cite{osc}.
Although the MSSM with explicit $R$-parity
violation could explain neutrino masses \cite{rp}, the virtue of having a DM 
candidate would then be lost in general.
Therefore, if one insists on having a DM candidate in the MSSM, one has
to include additional fields or particles, e.g. right-handed neutrinos,
in order to generate neutrino masses and oscillations.

One of the most celebrated ways to generate neutrino masses is
the seesaw mechanism~\cite{seesaw}, which is often considered 
to be the most natural and attractive.
The benefit and drawback of the original seesaw is that 
its scale is generically around the scale of grand unified theories
(GUTs) $\sim 10^{12-16}\gev$,
which is not accessible for direct phenomenological tests.
There have been many alternatives or modifications to
the original seesaw such that the seesaw scale can be as low as 
GeV or TeV which can be tested in current or future experiments.
Nevertheless, one big drawback of many of these models
is that the seesaw scale is put in rather ad-hoc by hand to be low.

One popular variant of these proposals is the inverse seesaw (ISS)
mechanism \cite{inverse},
which is an extension of the original seesaw model but with a 
much lower mass scale usually below several TeV. 
The inverse seesaw mechanism generates small neutrino masses
with rather large Yukawa couplings and 
violates lepton number mildly.
At such a low scale the model can be tested at Hadron colliders (the LHC and at 
future $100\tev$ colliders, e.g.~\cite{Das:2014jxa,Das:2015toa,100tev}) and future high energy lepton colliders 
(the Circular Electron Positron Collider, e.g.~\cite{Das:2012ze,Antusch:2015mia}, 
International Linear Collider, e.g.~\cite{Das:2012ze,Antusch:2015mia}, 
and the FCC-ee, e.g.~\cite{Blondel:2014bra}),
for an overview and comparison of the different collider possibilities, see,
e.g.~\cite{Antusch:2016ejd}.
Indirect effects can also be tested at low-energy flavor physics
experiments, e.g.~\cite{Atre:2009rg}, or in Higgs decays, e.g.~\cite{Arganda:2015naa}. 

There is a plethora of ISS models and 
not surprisingly we are by far not the first to discuss a supersymmetric
version. For the sake of brevity we give here a short overview of SUSY ISS
models only.
To our knowledge these models can be roughly categorized under one
of the four categories where each time the additional fields required in the
ISS have to be added:
(i) MSSM~\cite{Deppisch:2004fa,Arina:2008bb,Hirsch:2009ra,BhupalDev:2012ru}, 
(ii) MSSM/NMSSM with extended gauge 
symmetry~\cite{Khalil:2011tb,Hirsch:2011hg,Kang:2011wb}, 
(iii) NMSSM~\cite{Bazzocchi:2009kc, Das:2012ze}, and
(iv) supersymmetric Left-Right symmetry model~\cite{Dev:2009aw, An:2011uq}.
We briefly summarize these models as follows.

\begin{description}
\item[MSSM with additional gauge singlets]: \\
It has been pointed out in Ref.~\cite{Hirsch:2009ra} that 
by adding only one pair of gauge singlets 
$(S,N)$ to the MSSM it is sufficient 
to explain the neutrino data using an ISS.
One neutrino mass is generated at tree level 
while the other non-zero neutrino masses are 
generated by loop effects.
This is justifiable called the minimal
version of SUSY inverse seesaw model. In our approach we also
aim for minimality but we want to explain all neutrino masses 
at tree-level already which forces us to introduce two pairs of extra gauge singlets.
Of course it is also possible and popular to introduce three pairs of the 
extra gauge singlets with opposite lepton numbers, see, e.g.~\cite{Deppisch:2004fa, Arina:2008bb,BhupalDev:2012ru}.  

\item[The gauge extended SUSY]:\\ 
In this class of models, the seesaw mechanism is derived from
a symmetry breaking pattern of a $B-L$ extension of the MSSM.
The gauge group is $SU(3)_C\times SU(2)_L\times
U(1)_Y\times U(1)_{B-L}$, which gives rise to three SM singlets due to
the $U(1)_{B-L}$ anomaly cancellation conditions. These singlets can be
the right-hand neutrinos for the seesaw mechanism.
At the same time,  the lightest right-handed sneutrino could be 
the lightest SUSY particle (LSP)~\cite{Khalil:2011tb}.  
This is also attractive since it can be embedded into 
$SO(10)$ which was studied, e.g.\ in Ref.~\cite{Hirsch:2011hg},
where even an additional gauge factor $U(1)_R$ is introduced.
Interestingly, the sneutrino in this class of models can survive 
all the dark matter constraints in the inverse
seesaw extension but not in the linear seesaw.

\item[The NMSSM with an extra singlet sector]:\\ 
By adding an extra singlet sector to the NMSSM~\cite{Bazzocchi:2009kc}, 
tiny neutrino masses can be radiatively generated by 
the SUSY breaking parameters at very low scales similar
to \cite{Hirsch:2009ra} which we mentioned above.
In such a model, the sneutrino or the lightest neutralino could be 
the LSP~\cite{Bazzocchi:2009kc, Das:2012ze}.
In another NMSSM extension~\cite{Kang:2011wb} they connect
neutrino physics to asymmetric dark matter.

\item[$\boldsymbol{SU(2)_L\times SU(2)_R\times U(1)_{B-L}}$]:\\ 
In the last class of supersymmetric inverse seesaw implementations 
the MSSM gauge group is extended to a left-right symmetry $SU(2)_L\times
SU(2)_R\times U(1)_{B-L}$~\cite{Dev:2009aw, An:2011uq}. In these models 
the $B-L$ symmetry is broken at a low scale $\sim$TeV and the neutrino
masses are dynamically generated. Interestingly these models
can be embedded into $SO(10)$ models which reduces the effective number
of parameters making the model more predictive. 

\end{description}

In this paper we discuss a supersymmetric version of the ISS
where a $Z_6$ symmetry plays the role of lepton number which
is usually implemented as an approximate symmetry in ISS models.
Conventionally, $R$-parity is introduced to ensure proton
stability. This is not needed in our model since all $R$-parity violating
operators are already forbidden by the $Z_6$-symmetry.
The $Z_6$ is broken in the same way as the electroweak symmetry
in the MSSM and both scales are related to SUSY breaking.
Hence, in our model we have an intimate connection between the
seesaw scale and the TeV scale which gives a strong theoretical
motivation to have a low seesaw scale.

Our model is minimal not only with respect to symmetry extensions
but also with respect to the field content. 
Only five additional SM singlet fields are
introduced to the superpotential. The superfields $\hat{N}^c$ 
contain right-handed (RH) neutrinos and sneutrinos while the singlet
superfields $\hat{S}$ and $\hat{X}$ contain new singlet scalars and fermions.
Unlike the NMSSM, in which the new singlet superfield also couples to the 
two Higgs-doublet superfields, here the singlet superfields $\hat{S}$ 
and $\hat{X}$ only couple to the RH neutrino superfields $\hat{N}^c$ or to
themselves.
Our model hence would fall most closely under the first category of
an ISS extension of the MSSM since our additional symmetry is not gauged.

Since we gave up on some rather ad-hoc arguments about the scales
involved in the ISS we do not have to restrict ourselves to the the original
inverse seesaw mechanism with $M_S \ll M_D \ll \mu_{NS}$,
where $M_S$ is the singlets mass term,
$M_D$ the Dirac neutrino mass term and
$\mu_{NS}$ a supersymmetric mass term
respectively. 
To remind the reader the ISS neutrino mass matrix has
the structure
\begin{equation}
 M_\nu = \begin{pmatrix}
  0 & M_D & 0 \\
  M_D^T & 0 & \mu_{NS} \\
  0 & \mu_{NS}^T & M_S
 \end{pmatrix}  \label{Eq:numassIntro}
\end{equation}
in the basis $(\nu,\, N^c,\, S)^T$.
In our setup the original ordering of mass hierarchies
can be easily generalized to 
three different types of inverse seesaw mechanisms:
(i) $M_S \ll M_D \ll \mu_{NS}$ (ISS type I),
(ii) $M_S \approx M_D \ll \mu_{NS}$ (ISS type II), 
and 
(iii) $M_D \ll M_S \ll \mu_{NS}$ (ISS type III).
We investigate these three types of the ISS
and find that they can have very different phenomenology
which is expected since, for instance, the Yukawa couplings
turn out to be very different in size.

This work is organized as follows: First we describe the model in 
Sec.~\ref{sec:model} and in Sec.~\ref{sec:pheno} 
we discuss various phenomenological implications such as
the neutrino mass spectrum and 
mixing, neutrinoless double beta decay, and charged 
lepton-flavor violations.
We summarize and conclude in Sec.~\ref{sec:summary}.
In the appendix we have collected some explicit expressions
for mixing matrices which are too long for the main text.

\section{The model}
\label{sec:model}

In this section, we describe the model in detail, that is
the superpotential, the soft SUSY
breaking parameters, and the scalar potential.  The aim is to
construct a minimal supersymmetric inverse seesaw model. It is
minimal in the sense that we want to extend the MSSM with the least
possible extra fields and symmetries to get a viable inverse seesaw
mechanism to generate neutrino masses
at tree level which will be discussed in the next section.

\subsection{The superpotential}

\begin{table}
\centering
\begin{tabular}{lcccccccccc}
\toprule
Superfield & $\hat{Q}_i$ & $\hat{U}_i^c$ & $\hat{E}_i^c$ & $\hat{L}_i$ & $\hat{D}_i^c$ & $\hat{H}_u$ & $\hat{H}_d$ & $\hat{N}^c_\alpha$ & $\hat{S}_\alpha$ & $\hat{X}$ \\
\midrule 
$Z_6$ charge & 5 & 5 & 5 & 3 & 3 & 2 & 4 & 1 & 5 & 2 \\
\bottomrule
\end{tabular}
\caption{\label{tab:Model}
Superfield content of the model and charge assignment under the additional
discrete $Z_6$ symmetry. The new superfields compared to the MSSM, $\hat{N}^c$, $\hat{S}$ and $\hat{X}$,
are singlets under the Standard Model gauge group. The indices $i = 1$, 2, 3
and $\alpha = 1$, 2 are generation indices.
}
\end{table}

We impose a $Z_6$ symmetry on the superpotential under which the superfields transform as 
\begin{equation}
  \hat{\Phi} \to \hat{\Phi} \exp\left [ \text{i }  q \frac{2\pi}{6} \right ] \;,
\end{equation}
where $q$ runs from $0$ to $5$.  
The assignment of $q$ for the superfields in our model is listed in Table~\ref{tab:Model}.

This charge assignment is not unique but we have chosen the $Z_6$ charges such
that they are compatible with $SU(5)$ unification and such that we forbid the
$R$-parity violating operators of the MSSM. 
Because our superpotential does not conserve $U(1)$ lepton number, 
$R$-parity is not well defined.

The renormalizable superpotential compatible with the SM gauge symmetries 
and the $Z_6$ symmetry is then given by
\begin{equation}
  \mathcal{W} = \mathcal{W}_{\text{MSSM}} + \mathcal{W}_\nu \;,
\end{equation}
where
\begin{align}
\mathcal{W}_{\text{MSSM}} &= Y_u \, \hat{Q} \hat{H}_u \hat{U}^c - Y_d \, \hat{Q} \hat{H}_d \hat{D}^c - Y_e \, \hat{L} \hat{H}_d \hat{E}^c  + \mu_H \hat{H}_u \hat{H}_d \;, \\
\mathcal{W}_\nu       &= Y_\nu \, \hat{L} \hat{H}_u \hat{N}^c + \mu_{NS} \, \hat{N}^c \hat{S} + \frac{\lambda}{2} \, \hat{X} \, \hat{S}^2 + \frac{\kappa}{3} \, \hat{X}^3 \;,
\end{align}
and we have suppressed generation indices.
In our conventions, we label the superfields with a hat, the
fermionic components of the matter fields (including $\hat{N}^{c}$ and $\hat{S}$) without hat 
and their scalar components with a tilde. 
This is twisted for the Higgs
doublets and $\hat{X}$ (scalars without hat or tilde and fermions with a tilde).
Note that the superfields $\hat{N}^c$ will give rise to right-handed 
neutrinos and sneutrinos while the singlet superfields $\hat{S}$ and $\hat{X}$ will 
give rise to new singlet scalars and fermions. 

For the MSSM fields we assume the conventional number of generations.
To accommodate the neutrino masses and mixing at tree level we need at least two generations
of right-handed neutrino superfields $\hat{N}^c$ and two generations of additional
singlet superfields $\hat{S}$ for the realization of the ISS mechanism, see also \cite{Abada:2014vea}.
For the sake of simplicity this is what we assume throughout the rest of the paper. 
$\hat{X}$ gives rise to lepton number violation 
and its vacuum expectation value (vev) induces a Majorana mass term of $S$ 
as we will see in the next section.

\subsection{The soft SUSY breaking terms and the scalar potential}
\label{sec:SUSYbreaking}

The soft SUSY breaking terms can be grouped into the ordinary MSSM part
and additional terms
\begin{equation}
 - {\cal L}_{\rm soft} = - {\cal L}_{\rm soft, MSSM} - {\cal L}_{\rm soft, \nu} \;,
\end{equation}
where 
\begin{align}
- {\cal L}_{\rm soft, MSSM}  &= \frac{1}{2} M_1 \tilde{B} \tilde{B} 
              + \frac{1}{2} M_2 \tilde{W} \tilde{W}
              + \frac{1}{2} M_3 \tilde{g} \tilde{g} \nonumber \\
  &+ M_{\tilde{Q}}^2 \tilde{Q}^\dagger \tilde{Q} 
     +  M_{\tilde{U^c}}^2 \tilde{U^c}^\dagger \tilde{U_c} 
     +  M_{\tilde{D^c}}^2 \tilde{D^c}^\dagger \tilde{D_c}
     +  M_{\tilde{L}}^2 \tilde{L}^\dagger \tilde{L} 
     +  M_{\tilde{E^c}}^2 \tilde{E^c}^\dagger \tilde{E^c}\nonumber \\
  &  + M_{H_u}^2 {H_u}^\dagger H_u + M_{H_d}^2 {H_d}^\dagger H_d + (b_H H_u H_d +  \text{H.c.}) \nonumber \\
  &  + \left ( A_u \tilde{Q} H_u \tilde{U}^c - A_d \tilde{Q} H_d \tilde{D}^c 
     - A_e \tilde{L} H_d \tilde{E}^c + {\rm H.c.} \right)  \;, \\
- {\cal L}_{\rm soft, \nu}  &= M_{\tilde{N^c}}^2 \tilde{N}^{c \dagger} \tilde{N}^c
     +  M_{\tilde{S}}^2 \tilde{S}^\dagger \tilde{S}
     + M_{X}^2 X^\dagger X
     + (b_{NS} \tilde{N}^c \tilde{S} + \text{H.c.}) \nonumber \\
  &  + \left ( A_\nu \tilde{L} H_u \tilde{N}^c 
     + \frac{1}{2} A_{\lambda} X {\tilde{S}}^2
     + \frac{1}{3} A_{\kappa} X^3 + \text{H.c.} \right )\;, 
\end{align}
and where we have suppressed any gauge or generation indices.

To discuss the scalar potential we still have to add the $D$- and
$F$-terms.
Since the new states do not have gauge interactions we only have to
consider the $F$-terms for them which are given by 
$\sum_i | \partial W / \partial \phi_i |^2$, 
where $\phi_i$ is the scalar component of the superfields to be considered. 
The new part of the scalar potential then reads
\begin{align}
 V_{\text{new}} &= |Y_\nu \tilde{L} H_u + \mu_{NS} \tilde{S}|^2
     + |\mu_{NS} \tilde{N}^c +  \lambda \, X \, \tilde{S} |^2 
   \;+\;
  |- Y_e H_d \tilde{E}^c + Y_\nu H_u \tilde{N}^c |^2 
   + \left| \frac{1}{2}\lambda \, \tilde{S}^2 + \kappa \, X^2 \right|^2 \nonumber\\
    &+  M_{\tilde{N}^c}^2 \tilde{N^c}^\dagger \tilde{N^c}  
     +  M_{\tilde{S}}^2 \tilde{S}^\dagger \tilde{S}
     + M_{X}^2 X^\dagger X
     +  \left( b_{NS} \tilde{N}^c \tilde{S}  + \text{H.c.}
   \right )   \nonumber\\
     &+ \left( A_\nu \tilde{L} H_u \tilde{N}^c
     + \frac{1}{2} A_{\lambda} X \tilde{S}^{2}
     + \frac{1}{3} A_{\kappa} X^3  + \text{H.c.} \right )\;.
\end{align}
This is in general a very complicated potential since we have to consider
three generations of $\tilde{L}$ and $\tilde{E}^c$,
two generations of $\tilde{N}^c$, two generations of $\tilde{S}$, 
and one generation of $X$. 
In addition, this potential mixes with the 
conventional MSSM potential. Before we study this in more detail
we will restrict ourselves to the case of one generation of slepton doublets,
right-handed sneutrinos and scalar singlets each. 
We also assume that all couplings
and mass parameters are real which allows us to understand some essential
features and the rest is left to a future detailed numerical study of the model.

Since we do not want to introduce any additional source of electroweak
symmetry breaking we set $\langle \tilde{L} \rangle = 0$ and
$\langle \tilde{E}^c \rangle = 0$.
Keep in mind that choosing the appropriate parameters this is always possible,
since there is a $D$-term quartic in $\tilde{L}$ and a $D$-term
quartic in $\tilde{E}^c$ which dominates the potential for large field values and
the other parameters can be adjusted to allow only the trivial vacuum.

We define the vev of the relevant scalar fields as 
$\langle H_u^0 \rangle = v_u$, $\langle H_d^0 \rangle = v_d$,
$\langle \tilde{N}^c \rangle = v_N$, $\langle \tilde{S} \rangle = v_S$, 
and $\langle X \rangle = v_X$. 
The scalar potential is 
\begin{align}
 V_{\text{scalar}} 
&\supset
        (M_{H_u}^2 + \mu_H^2) v_u^2 + (M_{H_d}^2 + \mu_H^2) v_d^2 - 2 b_H v_u v_d 
      + \frac{1}{8} (g^2 + g'^2) (v_u^2 -v_d^2)^2 \nonumber\\
    &+ (\mu_{NS} v_S)^2 
      + (\mu_{NS} v_N + \lambda \, v_S \, v_X )^2 
      + ( Y_\nu v_u v_N )^2 + \left( \frac{1}{2} \lambda \, v^2_S +\kappa \, v_X^2 \right)^2 \nonumber\\
    &+  M_{\tilde{N}^c}^2 v_N^2
      +  M_{\tilde{S}}^2 v_S^2 + M_{X}^2 v_X^2 
      + 2 \left( b_{NS} v_N v_S
      + \frac{1}{2} A_{\lambda} v_X v_S^2
      + \frac{1}{3} A_{\kappa} v_X^3 \right ) 
      \nonumber\\
    &= 
     m_{H_u}^2 v_u^2 + Y_\nu^2 v_N^2 v_u^2 + m_{H_d}^2 v_d^2 - 2 b_H v_u v_d 
     + \frac{1}{8} (g^2 + g'^2) (v_u^2 -v_d^2)^2 \nonumber\\
    &+ m_S^2 v_S^2 + m_N^2 v_N^2 + M_{X}^2 v_X^2
      + v_N v_S \left( 2 b_{NS} + 2 \, \lambda \, \mu_{NS} \, v_X \right)
      + A_{\lambda} v_X v_S^2          
  \nonumber\\
    &+ \frac{2}{3} A_{\kappa} v_X^3 + \frac{1}{4} \lambda^2 \, v_S^4 
      + \kappa^2 \, v_X^4 +  (\lambda^2 +\lambda \, \kappa) \, v_S^2 v_X^2
       \;,
\end{align}
where we have set $m_{H_u}^2 = M_{H_u}^2 + \mu_H^2$,
$m_{H_d}^2 = M_{H_d}^2 + \mu_H^2$,  $m_S^2 = M_{\tilde{S}}^2 + \mu_{NS}^2$
and $m_N^2 = M_{\tilde{N}^c}^2 + \mu_{NS}^2$. The conventional
MSSM Higgs part was taken from \cite{Martin:1997ns}.

Now we are looking at the first derivatives to look for extrema of the potential
\begin{align}
\frac{\partial V_{\text{scalar}}}{\partial v_u} 
&= 
2 \left(m_{H_u}^2 + Y_\nu^2 v_N^2  \right) v_u - 2 b_H v_d + \frac{1}{2} (g^2 + g'^2) (v_u^3 - v_u v_d^2)
= 
0\,,  \label{eq:tadpole1}\\
\frac{\partial V_{\text{scalar}}}{\partial v_d} 
&= 
2 m_{H_d}^2 v_d - 2 b_H v_u + \frac{1}{2} (g^2 + g'^2) (v_d^3 - v_u^2 v_d)
= 
0\,,  \label{eq:tadpole2}\\
\frac{\partial V_{\text{scalar}}}{\partial v_S} 
&= 
2 m_S^2 v_S + v_N \left( 2 \, b_{NS} + 2  \,\lambda \, \mu_{NS} \, v_X \right) 
+ 2 A_{\lambda} v_X v_S + \lambda^2 v_S^3 + 2 (\lambda^2 +\lambda\kappa) v_S v_X^2 
= 
0 \;, \label{eq:tadpole3}\\
\frac{\partial V_{\text{scalar}}}{\partial v_N} 
&= 
2 (m_N^2  +  Y_\nu^2 v_u^2 ) v_N +  \left( 2 \, b_{NS} + 2 \, \lambda \, \mu_{NS} \, v_X \right) v_S
= 
0 \;, \label{eq:tadpole4}\\
\frac{\partial V_{\text{scalar}}}{\partial v_X} 
&= 
2 M_{X}^2 v_X + 2  \lambda \mu_{NS} v_N v_S 
+ A_{\lambda} v_S^2 + 2 A_{\kappa} v_X^2 + 4 \kappa^2  v_X^3 
+ 2  \left(\lambda^2 + \lambda\kappa \right) v_S^2 v_X
= 
0 \;. \label{eq:tadpole5}
\end{align}
Here, we would like to note several features of these tadpole conditions in
Eqs.~(\ref{eq:tadpole1})-(\ref{eq:tadpole5}). 
Once we switch off the vevs of the additional fields, i.e. $v_N = v_S = v_X = 0$, 
these tadpole conditions go back to the MSSM ones. 

The only viable solution from a phenomenological point of view of these tadpole
conditions is $v_N=v_S=0$ and $v_X \neq 0$. In particular we need
$v_X \neq 0$ to generate neutrino masses.
Its solution is 
\begin{align}
v_X 
= 
-\frac{A_{\kappa}}{4 \, \kappa^2}\pm\frac{\sqrt{A_{\kappa}^2 -8 \, \kappa^2 M_{X}^2}}{4 \, \kappa^2} \;.
\end{align}
This will be important later on and tells us that in our setup 
the neutrino mass scale is related to the scale of SUSY breaking
which is different from many ISS models 
where the right-handed neutrino masses are forbidden 
and the smallness of the fermionic singlet masses are put in by hand 
due to the approximate lepton number conservation. 
Therefore, our setup is minimal and we can derive all the masses
without any willful assumption. 

In principle one can now also discuss the second derivatives and study the
conditions for the potential to have a minimum but we do not find any
simple, important insights from there. 
In particular, the case above is a
simplified version of the model under study and the expressions get very lengthy
for a more realistic case. 
For the later discussion we just keep in mind that $X$ gets an
electroweak scale (= SUSY breaking scale) vev but $\tilde{S}$ and $\tilde{N}^c$ do not
receive a vev.

\section{Phenomenology}
\label{sec:pheno}

In this section, we discuss some phenomenological aspects of our model.  
Like in any supersymmetric model there is a huge amount of phenomenological 
aspects which could be discussed. 
In this work, we focus only on the features immediately related to neutrino
masses and mixing. That is the non-unitarity of the
Pontecorvo-Maki-Nakagawa-Sakata (PMNS) matrix, neutrinoless
double beta decay and charged-lepton flavor violations (cLFV).
Other aspects will be discussed in future publications.

\subsection{Leptonic masses and mixing}

We begin with the relevant Yukawa couplings and
mass terms relevant to the leptonic sector in the Lagrangian
\begin{align}
- {\cal L}_{\nu} &= - (Y_e)_{ij} L_i H_d E^c_j + (Y_\nu)_{i \alpha} L_i N^c_{\alpha} H_u 
  + (\mu_{NS})_{\alpha \beta} N^c_\alpha S_\beta + \frac{1}{2} \lambda_{\alpha \beta} S_\alpha S_\beta X
+ \text{H.c.},
\end{align}
where $i$, $j = 1$, $2$, $3$ and $\alpha$, $\beta = 1$, $2$.
We are working in a basis where the charged-lepton Yukawa couplings
are diagonal and
\begin{equation}
 m_l = y_l \, v \, \cos \beta \;,
\end{equation}
where $l = e$, $\mu$, $\tau$, $v = 174$~GeV and
$v \cos \beta = \langle H_d^0 \rangle$.

Since $X$ receives a vev we define the mass matrix
\begin{align}
 M_S &=  \lambda \, v_X \;.  
\end{align}
Note that $M_S$ is symmetric since it is a Majorana
mass matrix.

We also define a Dirac neutrino mass matrix for the neutrinos
\begin{equation}
 M_D = Y_\nu v \sin \beta \;,
\end{equation}
where $v \sin \beta = \langle H_u^0 \rangle$. 
Furthermore, the mixing term between $N^c$ and $S$ is $\mu_{NS}$.
Using these definitions it is easy to write down the full neutrino mass
matrix
\begin{equation}
 M_\nu = \begin{pmatrix}
  0 & M_D & 0 \\
  M_D^T & 0 & \mu_{NS} \\
  0 & \mu_{NS}^T & M_S
 \end{pmatrix}  \label{Eq:numass}
\end{equation}
in the basis $(\nu,\, N^c,\, S)^T$.
One will immediately recognise that this is the pattern
of a double or an inverse seesaw mechanism \cite{inverse}.
The double seesaw mechanism requires $M_S \gg \mu_{NS}$ whereas
the inverse seesaw mechanism requires $M_S \ll \mu_{NS}$. The latter seems
to be a more natural choice here since  $M_S$ is related to
a potentially small Yukawa coupling and a symmetry breaking.\footnote{
From that point of view an inverse seesaw mechanism is technically
natural a la 't-Hooft~\cite{tHooft:1979rat}.}

There is one important thing we would like to point out here.
In our model we have basically only one mass scale which
is the SUSY breaking scale (assuming that the $\mu$-parameters are
of the same order). This has to be seen in contrast to the conventional
seesaw models where there is another superheavy seesaw scale besides
the electroweak scale. Hence, in our model the question what triggers
these huge gap between the two scales simply does not occur.

The original definition of the inverse
seesaw mechanism implies $M_S \ll M_D \ll \mu_{NS}$. Here we
generalize this definition to realize three different types of the inverse
seesaw mechanisms according to the assumed hierarchies in the masses
\begin{itemize}
 \item[(i)] ISS type I: $M_S \ll M_D \ll \mu_{NS}$,
 \item[(ii)] ISS type II: $M_S \sim M_D \ll \mu_{NS}$,
 \item[(iii)] ISS type III: $M_D \ll M_S \ll \mu_{NS}$.
\end{itemize}
The different cases are in the end assumptions about the size of the
involved Yukawa couplings. Keep in mind that the electroweak and $Z_6$
symmetry breakings are related to soft SUSY breaking parameters and it
is plausible to assume that the vevs are similar in size. The sizes of
the Yukawa couplings are here not as well motivated and in the following
we discuss the three cases mentioned above.  Note that
these are simplified assumptions though. In reality, it could well be that
one generation looks
more like ISS type I while another generation behaves like type III.

One big advantage of this simplified assumption is that we can do a proper
expansion of the neutrino mass and mixing in terms of some expansion parameters,
which we discuss soon for the three cases mentioned above. 
Without loss of generality we also choose a basis where $\mu_{NS}$ is 
diagonal, which implies
in particular that $\mu_{NS}^T = \mu_{NS}$ from now on unless stated otherwise.

Before we go through the details of the different types we would like to
anticipate one common result: in all three cases, the leading order expression for
the light neutrino mass matrix is the same and given by 
\begin{equation} \label{eq:LightNuMasses}
 m_\nu = M_D \, \mu_{NS}^{-1} \, M_S \, \mu_{NS}^{-1} \, M_D^T \;,
\end{equation}
which is nothing else than the ordinary inverse seesaw formula
and the other heavier mass eigenstates have masses
of the order of $\mu_{NS} \sim$~TeV with small corrections.
From the above formula it is also obvious that the inverse seesaw mechanism
is in our model a direct consequence of the $Z_6$ breaking ($M_S \sim v_X$).

The above formula can be rewritten 
\begin{equation} 
 m_\nu = Y_\nu  \, \mu_{NS}^{-1} \, \lambda \, \mu_{NS}^{-1} \, Y_\nu^T v_u^2 v_X \sim Y_\nu \, \lambda \, Y_\nu^T  \, \mathcal{O}(\text{TeV}) \;,
\end{equation}
where we have used the working assumption that the dimensionful
quantities $v_u$, $v_X$ and  $\mu_{NS}$ are all of the same order.
The smallness of neutrino masses is hence completely given by
the moderate smallness of the Yukawa couplings $Y_\nu$ and $\lambda$.
Their size is related to the size of the respective expansion parameter
as we will discuss in the following for the three different ISS cases.

\subsubsection{ISS type I}

In this case we assume that $\mu_{NS}$
is $\mathcal{O}($TeV$)$, $M_D \sim \epsilon_{\text{I}} \, \mu_{NS}$ and
$M_S \sim \epsilon_{\text{I}}^2 \, \mu_{NS}$ where $\epsilon_{\text{I}}$
is the expansion parameter. We will quote the size of $\epsilon_{\text{I}}$ at
the end of this subsection after deriving the expression for the light
neutrino masses.

Note that we start with the product $M_\nu M_\nu^\dagger$
instead of $M_\nu$ alone.  
We diagonalize the matrix $M_\nu M_\nu^\dagger$ in two steps. 
First, we do a block rotation, $W$, to separate the light from the heavy states
sufficiently involving only small mixing angles.
Then we are left with another rotation $V$, which
acts upon the light and the heavy states
separately. In particular the rotation for the light states is the
PMNS matrix to a good approximation.  So our diagonalization condition
reads
\begin{equation}
 \label{eq:Diag_TypeI}
 \begin{split}
 U_{\text{I}} M_\nu M_\nu^\dagger U_{\text{I}}^\dagger &= V_{\text{I}} W_{\text{I}} M_\nu M_\nu^\dagger W_{\text{I}}^\dagger V_{\text{I}}^\dagger
   = V_{\text{I}} \begin{pmatrix}
         m_\nu m_\nu^\dagger & \mathcal{O}(\epsilon_{\text{I}}^7)  \\
         \mathcal{O}(\epsilon_{\text{I}}^7) & M_{R} M_{R}^\dagger
    \end{pmatrix} V_{\text{I}}^\dagger \\
   &=  \begin{pmatrix}
     U_{\text{PMNS}} & 0 \\
     0 & R_{\text{I}}
    \end{pmatrix}
    \begin{pmatrix}
         m_\nu m_\nu^\dagger & \mathcal{O}(\epsilon_{\text{I}}^7)  \\
         \mathcal{O}(\epsilon_{\text{I}}^7) & M_{R} M_{R}^\dagger
    \end{pmatrix}
    \begin{pmatrix}
     U_{\text{PMNS}}^\dagger & 0 \\
     0 & R_{\text{I}}^\dagger
    \end{pmatrix} \;,
  \end{split}  
\end{equation}
where $U_{\text{PMNS}}$ and $R_{\text{I}}$ diagonalize only the upper
3$\times$3 and the lower 4$\times$4 blocks, respectively.
As we will see very soon $m_\nu m_\nu^\dagger$ is
of $\mathcal{O}(\epsilon_{\text{I}}^8)$ and $M_{R} M_{R}^\dagger$ is of
$\mathcal{O}(1)$. Hence, the remaining off-diagonal elements of $\mathcal{O}(\epsilon_{\text{I}}^7)$
are negligible.

We present an explicit expression for $W_{\text{I}}$ and its elements
$w_{ij}$ in Appendix~\ref{sec:explicit}.
Here we just present $W_{\text{I}}$ and $U_{\text{I}}$ in terms of the leading order
in $\epsilon_{\text{I}}$
\begin{equation}
\label{eq:MixingI}
 W_{\text{I}} \sim \begin{pmatrix}
   1 & w_{12} \eta_{\text{I}}^3 & w_{13} \eta_{\text{I}} \\
   w_{21} \eta_{\text{I}}^3 & 1 & w_{23} \eta_{\text{I}}^8 \\
   w_{31} \eta_{\text{I}} & w_{32} \eta_{\text{I}}^4 & 1
  \end{pmatrix} \text{ and }
 U_{\text{I}} \sim \begin{pmatrix}
     U_{\text{PMNS}} & \eta_{\text{I}}^3 U_{\text{PMNS}} w_{12}  & \eta_{\text{I}}  U_{\text{PMNS}} w_{13} \\
     R_{\text{I}} \begin{pmatrix} w_{21} \eta_{\text{I}}^3\\w_{31} \eta_{\text{I}} \end{pmatrix} & \multicolumn{2}{c}{ R_{\text{I}} } 
    \end{pmatrix}
  \;.
\end{equation}
We have introduced here $\eta_{\text{I}} = 1$ which labels the 
order of the matrix elements in $\epsilon_{\text{I}}$. For instance,
we write $w_{12} \eta_{\text{I}}^3$ which states that the element
$w_{12}$ is $\mathcal{O}(\epsilon_{\text{I}}^3)$.

For the light and heavy mass matrices we only quote the leading and
next-to-leading order contributions
\begin{align}
 m_\nu m_\nu^\dagger &= 
     \eta_{\text{I}}^8 M_D  \mu_{NS}^{-1}  M_S  \mu_{NS}^{-1}  M_D^T M_D^*  (\mu_{NS}^*)^{-1} M_S^* (\mu_{NS}^*)^{-1} M_D^\dagger \nonumber\\
  & - \frac{1}{2} \eta_{\text{I}}^{10} M_D (\mu_{NS})^{-1}
          \Big( M_S (\mu_{NS})^{-1} M_D^T M_D^* (\mu_{NS}^*)^{-1}
            M_S^* (\mu_{NS}^*)^{-1} M_D^\dagger M_D (\mu_{NS})^{-1}
           \nonumber\\
  &  + 2 M_S (\mu_{NS})^{-1} M_D^T M_D^* (\mu_{NS}^*)^{-1} (\mu_{NS})^{-1}  M_D^T M_D^* (\mu_{NS}^*)^{-1} M_S^*  \nonumber\\
  &  +   (\mu_{NS}^*)^{-1} M_D^\dagger M_D (\mu_{NS})^{-1} M_S (\mu_{NS})^{-1} M_D^T M_D^* (\mu_{NS}^*)^{-1} M_S^* \Big)
    (\mu_{NS}^*)^{-1} M_D^\dagger \;, \\
 M_{R} M_{R}^\dagger &= \begin{pmatrix}
   \mu_{NS} \mu_{NS}^* + \eta_{\text{I}}^2 M_D^T   M_D^*  & \eta_{\text{I}}^2 \mu_{NS}   M_S^* \\
  \eta_{\text{I}}^2 M_S \mu_{NS}^*  &  \mu_{NS} \mu_{NS}^*
   + \tfrac{1}{2} \eta_{\text{I}}^2 ( \mu_{NS}  M_D^\dagger  M_D  \mu_{NS}^{-1}  + (\mu_{NS}^*)^{-1}  M_D^\dagger  M_D  \mu_{NS}^* )
   \end{pmatrix} ,
\end{align}
where we have quoted for convenience the orders
in $\epsilon_{\text{I}}$ explicitly using $\eta_{\text{I}}$. 

In our minimal setup $M_D$ is a 3$\times$2 matrix and therefore the lightest neutrino
is strictly massless due to rank considerations. Our neutrino mass scale is hence given by
$\sqrt{\Delta m_{32}^2} \approx 5 \cdot 10^{-2}$~eV and
$\epsilon_{\text{I}} \sim (0.01\text{ eV}/ \text{TeV})^{1/4} \sim 10^{-4}$.
This implies that $Y_\nu \sim 10^{-4}$ and $\lambda \sim 10^{-8}$.

\subsubsection{ISS type II}

In the ISS type II we have again that $\mu_{NS}$
is $\mathcal{O}($TeV$)$ but now
$M_D \sim M_S \sim \epsilon_{\text{II}} \, \mu_{NS}$. Our
diagonalization reads now
\begin{equation}
 \label{eq:Diag_TypeII}
 \begin{split}
 U_{\text{II}}  M_\nu M_\nu^\dagger U_{\text{II}}^\dagger &=
 V_{\text{II}} W_{\text{II}} M_\nu M_\nu^\dagger W_{\text{II}}^\dagger V_{\text{II}}^\dagger
   = V_{\text{II}} \begin{pmatrix}
         m_\nu m_\nu^\dagger & \mathcal{O}(\epsilon_{\text{II}}^5)  \\
         \mathcal{O}(\epsilon_{\text{II}}^5) & M_{R} M_{R}^\dagger
    \end{pmatrix} V_{\text{II}}^\dagger \\
   &=  \begin{pmatrix}
     U_{\text{PMNS}} & 0 \\
     0 & R_{\text{II}}
    \end{pmatrix}
    \begin{pmatrix}
         m_\nu m_\nu^\dagger & \mathcal{O}(\epsilon_{\text{II}}^5)  \\
         \mathcal{O}(\epsilon_{\text{II}}^5) & M_{R} M_{R}^\dagger
    \end{pmatrix}
    \begin{pmatrix}
     U_{\text{PMNS}}^\dagger & 0 \\
     0 & R_{\text{II}}^\dagger
    \end{pmatrix} \;.
  \end{split}  
\end{equation}
The neutrino mass matrices are
\begin{align}
 m_\nu m_\nu^\dagger &= \eta_{\text{II}}^6 M_D  \mu_{NS}^{-1}  M_S  \mu_{NS}^{-1}  M_D^T  M_D^*  (\mu_{NS}^*)^{-1}  
   M_S^*  (\mu_{NS}^*)^{-1}  M_D^\dagger \nonumber\\
   &- \frac{1}{2} \eta_{\text{II}}^8 M_D  \mu_{NS}^{-1}  M_S  \mu_{NS}^{-1}  M_D^T  M_D^*  (\mu_{NS}^*)^{-1} M_S^*  (\mu_{NS}^*)^{-1}  M_D^\dagger  M_D  \mu_{NS}^{-1}  (\mu_{NS}^*)^{-1}  M_D^\dagger \nonumber\\
   &- \eta_{\text{II}}^8 M_D  \mu_{NS}^{-1}  M_S  \mu_{NS}^{-1}  M_D^T  M_D^*  (\mu_{NS}^*)^{-1}  
   \mu_{NS}^{-1}  M_D^T  M_D^*  (\mu_{NS}^*)^{-1}  M_S^*  (\mu_{NS}^*)^{-1}  M_D^\dagger \nonumber\\
   &- \frac{1}{2} \eta_{\text{II}}^8 M_D  \mu_{NS}^{-1}  (\mu_{NS}^*)^{-1}  M_D^\dagger  M_D  \mu_{NS}^{-1}  M_S  
   \mu_{NS}^{-1}  M_D^T  M_D^*  (\mu_{NS}^*)^{-1}  M_S^*  (\mu_{NS}^*)^{-1}  M_D^\dagger \;,\\
 (M_{R} M_{R}^\dagger)_{11} &= \mu_{NS} \mu_{NS}^* + \eta_{\text{II}}^2 M_D^T   M_D^*\\
 (M_{R} M_{R}^\dagger)_{22} &= \mu_{NS} \mu_{NS}^* 
    + \eta_{\text{II}}^2 M_S  M_S^* 
    + 1/2 \eta_{\text{II}}^2  \mu_{NS}  M_D^\dagger  M_D  \mu_{NS}^{-1} 
    + 1/2 \eta_{\text{II}}^2 (\mu_{NS}^*)^{-1}  M_D^\dagger  M_D \mu_{NS}^* \\
 (M_{R} M_{R}^\dagger)_{12} &= \eta_{\text{II}}  \mu_{NS} M_S^* \\
 (M_{R} M_{R}^\dagger)_{21} &= \eta_{\text{II}}  M_S \mu_{NS}^*  \;.
\end{align}
For later reference we write
down explicitly $W_{\text{II}}$ and $U_{\text{II}}$ up to the leading orders in $\epsilon_{\text{II}}$
\begin{equation}
\label{eq:MixingII}
 W_{\text{II}} \sim \begin{pmatrix}
   1 & w_{12}\eta_{\text{II}}^2 & w_{13}\eta_{\text{II}} \\
     w_{21}\eta_{\text{II}}^2 & 1 & w_{23}\eta_{\text{II}}^5 \\
     w_{31}\eta_{\text{II}} & w_{32}\eta_{\text{II}}^3 & 1
  \end{pmatrix} \text{ and }
 U_{\text{II}} \sim \begin{pmatrix}
     U_{\text{PMNS}} & \eta_{\text{II}}^2 U_{\text{PMNS}} w_{12}  & \eta_{\text{II}}  U_{\text{PMNS}} w_{13} \\
     R_{\text{II}} \begin{pmatrix} w_{21} \eta_{\text{II}}^2\\ w_{31} \eta_{\text{II}} \end{pmatrix} & \multicolumn{2}{c}{ R_{\text{I}} } 
    \end{pmatrix}
  \;.
\end{equation}
We have introduced here $\eta_{\text{II}}$ to label the
order of the elements in $\epsilon_{\text{II}}$ for convenience
similar to ISS type I.
The explicit expression for $W_{\text{II}}$ can be found in Appendix~\ref{sec:explicit}.

For the expansion parameter $\epsilon_{\text{II}}$ in ISS type II 
we find $\epsilon_{\text{II}} \sim (0.01\text{ eV}/ \text{TeV})^{1/3} \sim 10^{-5}$
which is one order smaller than in ISS type I.
For the Yukawa couplings this implies $Y_\nu \sim \lambda \sim 10^{-5}$.

\subsubsection{ISS type III}

In the ISS type III we have again that $\mu_{NS}$
is $\mathcal{O}($TeV$)$ but now
$M_S \sim \epsilon_{\text{III}} \, \mu_{NS}$
and $M_D \sim \epsilon_{\text{III}}^2 \, \mu_{NS}$. Our
diagonalization reads here
\begin{equation}
 \label{eq:Diag_TypeIII}
 \begin{split}
  U_{\text{III}} M_\nu M_\nu^\dagger U_{\text{III}}^\dagger &=
  V_{\text{III}} W_{\text{III}} M_\nu M_\nu^\dagger W_{\text{III}}^\dagger V_{\text{III}}^\dagger
   = V_{\text{III}} \begin{pmatrix}
         m_\nu m_\nu^\dagger & \mathcal{O}(\epsilon_{\text{III}}^7)  \\
         \mathcal{O}(\epsilon_{\text{III}}^7) & M_{R} M_{R}^\dagger
    \end{pmatrix} V_{\text{III}}^\dagger \\
   &=  \begin{pmatrix}
     U_{\text{PMNS}} & 0 \\
     0 & R_{\text{III}}
    \end{pmatrix}
    \begin{pmatrix}
         m_\nu m_\nu^\dagger & \mathcal{O}(\epsilon_{\text{III}}^7)  \\
         \mathcal{O}(\epsilon_{\text{III}}^7) & M_{R} M_{R}^\dagger
    \end{pmatrix}
    \begin{pmatrix}
     U_{\text{PMNS}}^\dagger & 0 \\
     0 & R_{\text{III}}^\dagger
    \end{pmatrix} \;,
  \end{split}  
\end{equation}
The explicit expression for $W_{\text{III}}$ can be found in Appendix~\ref{sec:explicit}.
The neutrino mass matrices are
\begin{align}
 m_\nu m_\nu^\dagger &= \eta_{\text{III}}^{10} M_D \mu_{NS}^{-1} M_S \mu_{NS}^{-1} M_D^T M_D^* (\mu_{NS}^*)^{-1} M_S^* (\mu_{NS}^*)^{-1} M_D^\dagger +  \mathcal{O}(\eta_{\text{III}}^{14})   \;,\\
 M_{R} M_{R}^\dagger &= \begin{pmatrix}
    \mu_{NS} \mu_{NS}^* & \eta_{\text{III}} \mu_{NS} M_S^*  \\
   \eta_{\text{III}} M_S \mu_{NS}^* &  \mu_{NS} \mu_{NS}^* + \eta_{\text{III}}^2 M_S M_S^*
  \end{pmatrix} \;.
\end{align}
For later reference we write
down explicitly the leading orders of $W_{\text{III}}$ and $U_{\text{III}}$
\begin{multline}
\label{eq:MixingIII}
 W_{\text{III}} \sim \begin{pmatrix}
   1 & w_{12}\eta_{\text{III}}^3 & w_{13}\eta_{\text{III}}^2 \\
    w_{21}\eta_{\text{III}}^3 & 1 & \mathcal{O}(\eta_{\text{III}}^{13})  \\
    w_{31}\eta_{\text{III}}^2 & w_{32}\eta_{\text{III}}^5 & 1
  \end{pmatrix} \text{ and }\\
 U_{\text{III}} \sim \begin{pmatrix}
     U_{\text{PMNS}} & \eta_{\text{III}}^3 U_{\text{PMNS}} w_{12}  & \eta_{\text{III}}^2  U_{\text{PMNS}} w_{13} \\
     R_{\text{III}} \begin{pmatrix} w_{21} \eta_{\text{III}}^3\\w_{31} \eta_{\text{III}}^2 \end{pmatrix} & \multicolumn{2}{c}{ R_{\text{III}} } 
    \end{pmatrix}
  \;.
\end{multline}
We have introduced here $\eta_{\text{III}}$ to label the
order of the elements in $\epsilon_{\text{III}}$ for convenience
similar to ISS type I.
The explicit expression for $W_{\text{III}}$ can be found in Appendix~\ref{sec:explicit}.

For the expansion parameter in ISS type III we find
$\epsilon_{\text{III}} \sim (0.01\text{ eV}/ \text{TeV})^{1/5} \sim 10^{-3}$.
ISS type III exhibits hence the mildest hierarchies
and it
has the smallest neutrino Yukawa couplings, $Y_\nu \sim 10^{-6}$,
and the largest singlet Yukawa coupling, $\lambda \sim 10^{-3}$.

It is also remarkable that in all three cases the leading order formulas
for the light and the heavy neutrino masses are the same but there are
differences in the next-to-leading order terms which might potentially help
to disentangle the three cases in precision measurements in the future.

\subsubsection{Non-unitarity of the mixing matrix}
\label{sec:nonUnitarity}

At this point we would like to comment on the non-unitarity of
the PMNS matrix obtained in our formalism. 
Only the full 7$\times$7 mixing matrix will be unitary while
any given sub-matrix of this matrix does not have to be unitary.
Let us illustrate this first with the ISS type I as an example
\begin{align}
 1 &= U_{\text{I}} U_{\text{I}}^\dagger = \begin{pmatrix}
     U_{\text{PMNS}} & 0 \\
     0 & R_{\text{I}}
    \end{pmatrix}
    W_{\text{I}} W_{\text{I}}^\dagger
    \begin{pmatrix}
     U_{\text{PMNS}}^\dagger & 0 \\
     0 & R_{\text{I}}^\dagger
    \end{pmatrix} \nonumber \\
  \sim&  \begin{pmatrix}
     U_{\text{PMNS}} & \eta_{\text{I}}^3 U_{\text{PMNS}} w_{12}  & \eta_{\text{I}}  U_{\text{PMNS}} w_{13} \\
     R_{\text{I}} \begin{pmatrix} w_{21} \eta_{\text{I}}^3\\w_{31} \eta_{\text{I}} \end{pmatrix} & \multicolumn{2}{c}{ R_{\text{I}} } 
    \end{pmatrix}  \begin{pmatrix}
     U_{\text{PMNS}}
      & \eta_{\text{I}}^3 U_{\text{PMNS}} w_{12}  & \eta_{\text{I}}  U_{\text{PMNS}} w_{13} \\
     R_{\text{I}} \begin{pmatrix} w_{21} \eta_{\text{I}}^3\\w_{31} \eta_{\text{I}} \end{pmatrix} & \multicolumn{2}{c}{ R_{\text{I}} } 
    \end{pmatrix}^\dagger
  \;.
\end{align}
For simplicity, we consider now only the first 3$\times$3 block
up to $\mathcal{O}(\epsilon_{\text{I}}^2)$
which we are interested in
\begin{align}
 1 &\approx U_{\text{PMNS}} U_{\text{PMNS}}^\dagger + \eta_{\text{I}}^2 U_{\text{PMNS}} w_{13} w_{13}^\dagger U_{\text{PMNS}}^\dagger \;.
\end{align}
Note that expanding in $\epsilon_{\text{I}}$
or $\eta_{\text{I}}$ gives here the same results since they
always appear together at the same order. Of course, technically
speaking we have to expand in $\epsilon_{\text{I}}$ since this is
the small parameter while $\eta_{\text{I}}$ is only a bookkeeping parameter
equal to one (and not small).
After multiplying this equation from left with $U_{\text{PMNS}}^{-1}$,
from right with $(U_{\text{PMNS}}^\dagger)^{-1}$ and inverting the whole
equation we find
\begin{align}
 U_{\text{PMNS}}^\dagger U_{\text{PMNS}} \approx (1 + \eta_{\text{I}}^2 w_{13} w_{13}^\dagger)^{-1} \;,
\end{align}
so that the deviation from unitarity is of
$\mathcal{O}(\epsilon_{\text{I}}^2)$ = $\mathcal{O}(10^{-8})$, which is 
much smaller than current constraints, see, for example,~\cite{Fernandez-Martinez:2016lgt},
but might be relevant in the future.

For the other two ISS types we find even smaller deviations
from unitarity of $\mathcal{O}(\epsilon_{\text{II}}^2)$ = $\mathcal{O}(10^{-10})$
and $\mathcal{O}(\epsilon_{\text{III}}^4)$ = $\mathcal{O}(10^{-12})$,
respectively.

\subsubsection{The Yukawa couplings}

Although at this point we do not need it explicitly we 
derive some expressions for the  Dirac neutrino Yukawa coupling
constants in terms of the Casas-Ibarra parameterization~\cite{Casas:2001sr}.
The advantage is that after fixing unknown parameters we can
immediately calculate the Yukawa matrix such that neutrino oscillation data
is correctly reproduced in our model.
From the neutrino mass matrix
Eq.~\eqref{eq:LightNuMasses}, the leading contribution to active neutrino masses
is obtained as 
\begin{align}
m_i \equiv 
U_{\rm PMNS}^\dagger \,
m_\nu  \,
U_{\rm PMNS}^\ast
= 
U_{\rm PMNS}^\dagger
M_D \mu_{NS}^{-1} M_S (\mu_{NS}^{-1})^T M_D^T 
U_{\rm PMNS}^\ast\,,\label{Eq:ISSpre}
\end{align}
where $m_i$ is the diagonal mass matrix of light, active neutrino states,
$m_i = {\rm diag}(m_1\,,m_2\,,m_3)$. Note that we have transposed
here the second $\mu_{NS}^{-1}$ for later convenience, which we have not
before since we are working in a basis where $\mu_{NS}$ is diagonal.

Since $M_S$ is not diagonal in general, we first need to diagonalize 
this matrix by a unitary matrix $V_S$, $M_S^d \equiv V_S M_S V_S^T$.
We can use this in Eq.~\eqref{Eq:ISSpre} and find 
\begin{align}
\sqrt{m_i} \, \sqrt{m_i} =
U_{\rm PMNS}^\dagger
M_D \mu_{NS}^{-1} V_S^{\dagger} \, \sqrt{M_S^d} \, \sqrt{M_S^d} \, V_S^* (\mu_{NS}^{-1})^T M_D^T 
U_{\rm PMNS}^\ast\,,\label{Eq:ISSaft}
\end{align}
from where we can easily derive the leading order expression for the
neutrino Yukawa couplings
\begin{align}
Y_\nu = \frac{\text{i}}{v_u} U_{\rm PMNS} \sqrt{m_i} \, \Omega \, \left( \sqrt{ M_S^d } \right)^{-1} V_S \, \mu_{NS}\,,
\end{align}
where $\Omega$ is an arbitrary, orthogonal, complex matrix 
parameterized by 
\begin{align}
\Omega^{\rm NH} = 
\begin{pmatrix}
0 &0\\
\cos \omega &\sin \omega\\
-\xi \sin \omega &\xi \cos \omega
\end{pmatrix}\,,
\hspace{1cm}
\Omega^{\rm IH} = 
\begin{pmatrix}
\cos \omega &\sin \omega\\
-\xi \sin \omega &\xi \cos \omega\\
0 &0
\end{pmatrix}\,,
\end{align}
with $\omega$ being a complex parameter and  $\xi = \pm 1$ corresponding to a
parity degree of freedom.
Here NH denotes normal neutrino mass hierarchy and IH denotes inverted
neutrino mass hierarchy.

\subsection{Neutrinoless double beta decay}

Once massive Majorana neutrinos are implemented into the SM, 
global lepton number symmetry is broken by two units and an
interesting phenomenom 
called neutrinoless double beta ($0 \nu \beta \beta$) decay
can occur, for a recent review see~\cite{Pas:2015eia}. 
The rate of $0 \nu \beta \beta$ decay is proportional to the modulus square of 
the effective mass, $m_{\rm eff}$, which is gradually constrained by several
experiments. The most stringent bound so far, $|m_{\rm eff}| < (61 \mathchar`- 161)~ {\rm meV}$,
is from the search for $0 \nu \beta \beta$
decay of ${}^{136}{\rm Xe}$ by the KamLAND-Zen collaboration~\cite{KamLAND-Zen:2016pfg}.

When we introduce only three massive Majorana neutrinos, $m_{\rm eff}$ can be expressed as 
\begin{eqnarray}
m_{\rm eff}^{\rm active} = \sum_{i=1}^3 (U_{\text{PMNS}})_{ei}^2 m_i\,,\label{Eq:meff_SM}
\end{eqnarray}
where $m_i$ are the mass eigenvalues of the neutrinos and $m_1$ ($m_3$) is exactly
equal to zero in our model for the NH (IH) case. 
When we take the active neutrino mass and the mixing angles given
in~\cite{Gonzalez-Garcia:2014bfa} we obtain that 
$|m_{\rm eff}|$ is $\mathcal{O}(1)$ and $\mathcal{O}(10)$ meV in the NH and IH cases, respectively. 
In addition to this standard contribution coming from the active neutrinos, 
one can get some other contributions in extensions of the minimal model
with three light active neutrinos only. 

Especially, as we have mentioned before, 
there is no $R$-parity in our model, 
and one might think of some contributions from the exchange of SUSY particles.
The SUSY contributions of $0 \nu \beta \beta$ decay are induced 
by the $R$-parity violating $\hat{L} \hat{Q} \hat{D}^c$ interaction of the first
generation~\cite{Mohapatra:1986su}.
However, such a term is forbidden due to the $Z_6$ symmetry 
imposed in our model. 
It means that we do not have any SUSY contribution to 
$0 \nu \beta \beta$ decay. 

As a result, we can focus on the non-SUSY contributions of the model. 
The contributions can be parameterized as 
\begin{eqnarray}
m_{\rm eff}^{\rm new} = \sum_{i=4}^7 (U_{ei})^2 m_i f_\beta (m_i)\,,
\end{eqnarray}
where $f_\beta(x)$ denotes the suppression factor of the nuclear matrix element 
when the mass scale $x$ is larger than a typical scale 
$\mathcal{O} (100~{\rm MeV})$. 
Since the typical mass scales for additional gauge singlet 
fermions is $\mu_{NS}$ as we discussed above, 
we simply replace $x$ by $x = \mu_{NS}$ 
and treat $\mu_{NS}$ like a number for simplicity
throughout the rest of this section.
In the current analysis, we adopt the expression 
\begin{eqnarray}
f_\beta (\mu_{NS}) = \frac{\langle p^2 \rangle}{\mu_{NS}^2 + \langle p^2 \rangle}\,,
\end{eqnarray}
with the typical momentum in the matrix element 
$\langle p^2 \rangle \simeq (200~{\rm MeV})^2$~\cite{Faessler:2014kka}. 
As the typical mass scale of the heavy neutrinos is of $\mathcal{O} ({\rm TeV})$, 
their contribution is given by~\cite{Abada:2014vea,Haba:2016lxc} 
\begin{eqnarray}
m_{\rm eff}^{\rm new} \eqn{\simeq} 
\sum_{i=4}^7 (U_{ei})^2 \frac{\langle p^2 \rangle}{\mu_{NS}^2} m_i\notag\\
\eqn{=} 
\langle p^2 \rangle
\left[ 
- (U_{e4})^2 
\frac{|m_4|}{\mu_{NS}^2}
+ (U_{e5})^2 
\frac{|m_5|}{\mu_{NS}^2}
- (U_{e6})^2 
\frac{|m_6|}{\mu_{NS}^2}
+ (U_{e7})^2 
\frac{|m_7|}{\mu_{NS}^2}
\right]
\,. 
\end{eqnarray}
Due to the particular structure of the mass matrix
there are always two mass eigenstates with almost the
same mass but opposite sign as suggested in the above formula.
To be more precise all the absolute values of the heavier masses are given
at the leading order by $\mu_{NS}$. At this order the cancellation is exact 
since also $(U_{e4})^2 = (U_{e5})^2$ and $(U_{e6})^2 = (U_{e7})^2$.
Nevertheless, this cancellation is not exact to all orders 
and the first non-vanishing order in ISS type I is obtained as 
\begin{eqnarray}
m_{\rm eff}^{\rm new} \eqn{\simeq} 
\sum_{i=4}^7 
(U_{ei})^2 \frac{\langle p^2 \rangle}{\mu_{NS}^2} M_S \lesssim 
\epsilon_\text{I}^4 \cdot \left( 8 \times 10^{7} \, {\rm meV} \right) \cdot \left( \frac{{\rm TeV}}{\mu_{NS}} \right)
\approx 8 \times 10^{-9} \, {\rm meV}  \cdot \left( \frac{{\rm TeV}}{\mu_{NS}} \right)
\,,
\end{eqnarray}
which is negligibly small compared to the contribution from the
light active neutrinos. In ISS type II and III the contributions
are even smaller as can be easily checked.

\subsection{Charged lepton flavor violation}

In the SM, cLFV is not allowed on the perturbative level,
but this will immediately change once neutrino masses are introduced. 
In the following we discuss some estimates for cLFV in our model.

\subsubsection{The non-SUSY part}

Let us begin with the discussion on the non-SUSY part which corresponds
to sending the SUSY breaking scale to infinity and there are no contributions
from the SUSY partners to the process $\mu \to e \gamma$ for instance.
To estimate the contributions of these processes we refer to an early
calculation by Cheng and Li \cite{Cheng:1980tp}, see also \cite{Petcov:1976ff},
but adapt their notation to
our conventions. We quote for simplicity the formulas for
$\mu \to e \gamma$ only. 
The expressions can be  straight-forwardly extended to other processes.
Each neutrino-like mass eigenstate contributes to the amplitude
\begin{equation}
 A_i = \frac{G_F}{\sqrt{2}} \frac{e \, m_\mu}{32 \, \pi^2} \,  U_{i e} U^*_{i \mu} \, F(m_i^2 /M_W^2) \;, 
\end{equation}
where $i= 1, \ldots, 7$ and
\begin{equation}
 F(x) = \frac{10 - 43 x + 78 x^2 - 49 x^3 + 4 x^4 + 18 x^3 \log x}{3 (x-1)^4}
\end{equation}
for $x>0$ and $x \neq 1$. For $x \ll 1$ this simplifies to
\begin{equation}
 F(x) = \frac{10}{3} - x + \mathcal{O}(x^2)
\label{103}
\end{equation}
and for $x \gg 1$
\begin{equation}
 F(x) = \frac{4}{3} - \frac{1}{x} \left(11 + 6 \log \frac{1}{x} \right) + \mathcal{O}(x^{-2}) \;.
\end{equation}

The physical branching ratio (BR) is $\propto |\sum_i A_i|^2$.
So we shall first identify the largest amplitude $A_i$ to get a
feeling for the maximal BR we can expect.
The neutrino-like
mass eigenstates are either much lighter or much heavier than
the $W$-boson, $x\ll1$ or $1/x \ll 1$. 
Therefore, the dominant
contribution is coming from the constant term of $F(x)$
and we consider the two cases separately. 
Let us begin
with the light states, $i = 1$, 2, 3. If there would be
only three light states which do not mix with any other states
we would find $\sum_{i=1}^3 U_{i e} U^*_{i \mu} = 0$ due to
the unitarity of the PMNS matrix. 
Therefore, the leading term contributions proportional to the
constant term in Eq.~(\ref{103}) all cancel out. The next leading term
in Eq.~(\ref{103}) would be $x = m_\nu^2/ M_W^2 \sim 10^{-20}$,
which is negligible compared with the incomplete unitarity that we discuss
now.
The unitarity is only complete when summed over all $i=1 - 7$, see the discussion
in Section~\ref{sec:nonUnitarity}, and therefore
\begin{equation}
 \sum_{i=1}^3 U_{i e} U^*_{i \mu} =
  \begin{cases}
    \mathcal{O}(\epsilon_{\text{I}}^2) = \mathcal{O}(10^{-8}) & \text{for ISS type I,} \\
    \mathcal{O}(\epsilon_{\text{II}}^2) = \mathcal{O}(10^{-10}) & \text{for ISS type II,} \\
    \mathcal{O}(\epsilon_{\text{III}}^4) = \mathcal{O}(10^{-12}) & \text{for ISS type III.} 
  \end{cases}
\end{equation}
Since $U$ as a $7\times7$ matrix is unitary this non-vanishing
has to be compensated by the heavy states such that we find as well
\begin{equation}
 \sum_{i=4}^7 U_{i e} U^*_{i \mu} =
  \begin{cases}
    \mathcal{O}(\epsilon_{\text{I}}^2) = \mathcal{O}(10^{-8}) & \text{for ISS type I,} \\
    \mathcal{O}(\epsilon_{\text{II}}^2) = \mathcal{O}(10^{-10}) & \text{for ISS type II,} \\
    \mathcal{O}(\epsilon_{\text{III}}^4) = \mathcal{O}(10^{-12}) & \text{for ISS type III.} 
  \end{cases}
\end{equation}
The branching ratio is defined with respect to the
width of the muon $\Gamma(\mu \to e \nu \bar\nu) = m_\mu^5 G_F^2/192 \pi^3$
such that we find
\begin{align}
 \text{BR}(\mu\to e \gamma) &= \frac{48 \, \pi^2 |\sum_i A_i|^2}{m_\mu^2 G_F^2} = 
  \frac{3 \alpha}{32 \pi} \left| \sum_i U_{i e} U^*_{i \mu} \, F(m_i^2 /M_W^2) \right|^2 \nonumber\\
   & = \begin{cases}
    \mathcal{O}(10^{-20}) & \text{for ISS type I,} \\
    \mathcal{O}(10^{-24}) & \text{for ISS type II,} \\
    \mathcal{O}(10^{-28}) & \text{for ISS type III,} 
  \end{cases}
\end{align}
which are all far below the current bound
$\text{BR}(\mu\to e \gamma) < 4.2 \times 10^{-13}$
at 90\% confidence level of the MEG experiment
\cite{TheMEG:2016wtm}.
The branching ratios for other
cLFV processes are similarly
suppressed but their bounds are generally weaker.

\subsubsection{The SUSY part}

\begin{figure}
 \centering
 \hspace{7mm}
 \includegraphics{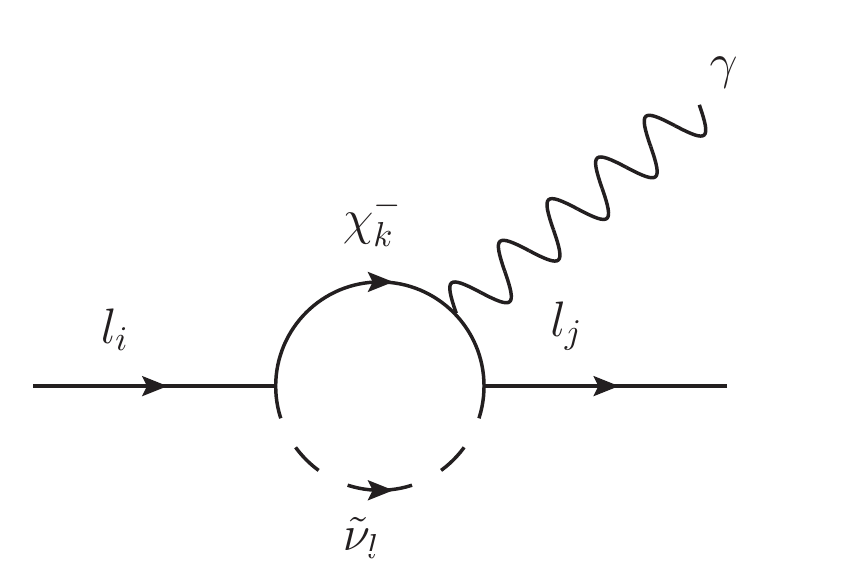}
 \caption{Feynman diagram for charged lepton flavor violation
 which changes compared to the case of the MSSM extended by right-handed neutrinos.
 \label{fig:SUSY_cLFV}}
\end{figure}

There are also contributions to cLFV 
from loops involving supersymmetric partners \cite{Hisano:1995cp}.
While the pieces involving the charged sleptons do not change, there
are major changes for the contributions involving scalar partners of
the neutrinos (of both chiralities) and the singlets, cf.~Fig.~\ref{fig:SUSY_cLFV}. 
Importantly, $X$ receives a vev which induces a mass splitting
for the CP-even and CP-odd components of the sneutrinos and
we define
\begin{align}
 \tilde{\nu}_L &= \frac{1}{\sqrt{2}} \left( \phi_L + \text{i } \sigma_L \right) \;,\\
 \tilde{N} &=  \frac{1}{\sqrt{2}} \left( \phi_R + \text{i } \sigma_R \right) \;,\\
 \tilde{S} &=  \frac{1}{\sqrt{2}} \left( \phi_S + \text{i } \sigma_S \right) \;.
\end{align}

Note though that the experimental bounds on cLFV from the SUSY contributions 
can always be satisfied by making the SUSY states heavy enough. 
And at this point we have no constraint on the SUSY
scale. In the future we plan to put our model into the \texttt{SARAH}
package \cite{Staub:2008uz} such that we can include additional constraints
and give more quantitative statements.
Note that if there are sources of CP violation CP-even and CP-odd
scalars can mix with each other which we neglect here.

In fact, we used the \texttt{SARAH} code 
to derive the following
expressions for the scalar mass matrices.
In the basis $(\phi_L,\phi_N,\phi_S)$ the mass matrix 
for the CP-even sneutrinos reads
\begin{equation} 
m^2_{\tilde{\nu}^R} = \left( 
\begin{array}{ccc}
m_{\phi_{L}\phi_{L}} &m^T_{\phi_{R}\phi_{L}} & v_u {\Re\Big({Y_{\nu}^{T}  \mu_{NS}^*}\Big)} \\ 
m_{\phi_{L}\phi_{R}} &m_{\phi_{R}\phi_{R}} & v_{X} {\Re\Big({\mu_{NS}  \lambda^*}\Big)}  + {\Re\Big(b_{NS}\Big)}\\ 
v_u {\Re\Big({\mu_{NS}^{T}  Y_{\nu}^*}\Big)}  & v_{X} {\Re\Big({\lambda  \mu_{NS}^{\dagger}}\Big)}  + {\Re\Big(b_{NS}^{T}\Big)} &m_{\phi_{S}\phi_{S}}\end{array} 
\right) \;,
 \end{equation} 
 where
\begin{align} 
m_{\phi_{L}\phi_{L}} &=   v_{u}^{2} {\Re\Big({Y_{\nu}^{T}  Y_{\nu}^*}\Big)}  + 2 \,{\Re\Big(M_{\tilde{L}}^2\Big)} + \frac{1}{2} M_Z^2 \cos (2 \beta) \;,\\ 
m_{\phi_{L}\phi_{R}} &= -   v_d \, {\Re\Big(Y_{\nu} \mu_H^* \Big)} + v_u \, {\Re\Big(A_{\nu}\Big)}  \;,\\ 
m_{\phi_{R}\phi_{R}} &= {\Re\Big(M_{\tilde{N}^c}^2\Big)}  +  \, {\Re\Big({\mu_{NS}  \mu_{NS}^{\dagger}}} \Big) +  v_{u}^{2} \, {\Re\Big({Y_{\nu}  Y_{\nu}^{\dagger}}\Big)}  \;,\\ 
m_{\phi_{S}\phi_{S}} &=    \, {\Re\Big(M_{\tilde{S}}^2\Big)}  +  \, {\Re\Big({\mu_{NS}^{T}  \mu_{NS}^*} } \Big) + v_{X} \Big(  \, {\Re\Big(A_{\lambda}\Big)}  + v_{X} \Big( {\Re\Big(\lambda \, \kappa^* \Big)}  +  {\Re\Big({\lambda  \, \lambda^*}\Big)} \Big)\Big) \;.
\end{align} 
This matrix is diagonalized by \(Z^R\): 
\begin{equation} 
Z^R m^2_{\tilde{\nu}^R} Z^{R,\dagger} = d^2_{\tilde{\nu}^R} \;,
\end{equation} 
with 
\begin{align} 
\begin{pmatrix}
 \phi_L \\
 \phi_N \\
 \phi_S
\end{pmatrix} = Z^{R,\dagger} \tilde{\nu}^{R}.
\end{align}

In the basis $(\sigma_L,\sigma_N,\sigma_S)$ the mass matrix
for the CP-odd sneutrinos reads
\begin{equation} 
m^2_{\nu^I} = \left( 
\begin{array}{ccc}
m_{\sigma_{L}\sigma_{L}} &m^T_{\sigma_{R}\sigma_{L}} & v_u {\Re\Big({Y_{\nu}^{T}  \mu_{NS}^*}\Big)} \\ 
m_{\sigma_{L}\sigma_{R}} &m_{\sigma_{R}\sigma_{R}} &-  v_{X} {\Re\Big({\mu_{NS}  \lambda^*}\Big)}  + {\Re\Big(b_{NS}\Big)}\\ 
 v_u {\Re\Big({\mu_{NS}^{T}  Y_{\nu}^*}\Big)}  &-  v_{X} {\Re\Big({\lambda  \mu_{NS}^{\dagger}}\Big)}  + {\Re\Big(b_{NS}^{T}\Big)} &m_{\sigma_{S}\sigma_{S}}\end{array} 
\right)  \;,
 \end{equation} 
 where
\begin{align} 
m_{\sigma_{L}\sigma_{L}} &=  v_{u}^{2} {\Re\Big({Y_{\nu}^{T}  Y_{\nu}^*}\Big)}  +  \,{\Re\Big(M_{\tilde{L}}^2\Big)}  + \frac{1}{2} M_Z^2 \cos (2 \beta) \;, \\ 
m_{\sigma_{L}\sigma_{R}} &= - v_d \, {\Re\Big(Y_{\nu} \mu_H^* \Big)} + v_u \, {\Re\Big(A_{\nu}\Big)}  \;,\\ 
m_{\sigma_{R}\sigma_{R}} &=  \, {\Re\Big(M_{\tilde{N}^c}^2\Big)}  +  \, {\Re\Big({\mu_{NS}  \mu_{NS}^{\dagger}}} \Big) +  v_{u}^{2} \, {\Re\Big({Y_{\nu}  Y_{\nu}^{\dagger}}\Big)}  \;,\\ 
m_{\sigma_{S}\sigma_{S}} &=  \, {\Re\Big(M_{\tilde{S}}^2\Big)}  +  \, {\Re\Big({\mu_{NS}^{T}  \mu_{NS}^*} } \Big) - v_{X} \Big( {\Re\Big(A_{\lambda}\Big)}  + v_{X} \Big( {\Re\Big(\lambda \, \kappa^* \Big)}  -  {\Re\Big({\lambda  \, \lambda^*}\Big)} \Big)\Big) \;.
\end{align}
This matrix is diagonalized by \(Z^I\): 
\begin{equation} 
Z^I m^2_{\tilde{\nu}^I} Z^{I,\dagger} = d^2_{\tilde{\nu}^I} \;,
\end{equation} 
with 
\begin{align} 
\begin{pmatrix}
 \sigma_L \\
 \sigma_N \\
 \sigma_S
\end{pmatrix} = Z^{I,\dagger} \tilde{\nu}^I \;.
\end{align}
Note that the difference between the two mass matrices is proportional
to the vevs $v_N$ and $v_S$ as expected.

To make the computation easier we follow a similar approach
as described in \cite{Cheung:2001sb}. Instead of treating CP-even and CP-odd
scalars separately we define a larger set of sneutrino states
\begin{equation}
  \tilde{\nu} = \begin{pmatrix}
                 \tilde{\nu}^R\\
                 \tilde{\nu}^I
                \end{pmatrix} \;,
\end{equation}
which has a block diagonal mass matrix
\begin{equation}
 m^2_{\tilde{\nu}} = \begin{pmatrix}
                       m^2_{\tilde{\nu}^R} & 0 \\
                       0 & m^2_{\tilde{\nu}^I}
                     \end{pmatrix} \;,
\end{equation}
which is diagonalized by $Z^{\tilde{\nu}}$: 
\begin{equation} 
Z^{\tilde{\nu}} m^2_{\tilde{\nu}} Z^{\tilde{\nu},\dagger} = d^2_{\tilde{\nu}} \;,
\end{equation} 
which is of course also block diagonal and unitary.

The vertices of the sneutrinos coupling to charginos and
charged leptons, which are the relevant vertices here,
can be easily reconciled from the MSSM vertices extended
by right-handed neutrinos. Due to the normalisation of the
fields, they get rescaled by a factor of $1/\sqrt{2}$ and the
couplings to the CP-odd scalars receive an additional factor of
$\text{i}$ for the incoming vertex and a factor of $-\text{i}$
for the outgoing vertex, cf.~Fig.~\ref{fig:SUSY_cLFV}. 
Therefore, the contributions from the CP-even and CP-odd scalars 
can be added up.
In the limit of vanishing $v_X$ we 
obtain the correct result
as if there were only seven complex sneutrinos.

At this point we will not go into any more details for the
full computation. There are excellent and detailed calculations
for cLFV in supersymmetric inverse seesaw
models in the literature, e.g.~\cite{Abada:2014kba}. Instead, we want to discuss
a bit more the qualitative features of the sneutrino mass matrices.

As we have discussed before in Section~\ref{sec:SUSYbreaking}
the vev $v_X$ is expected to be of the same order as the soft SUSY breaking
parameters.
Furthermore, it is a generic assumption that the soft trilinear couplings
are proportional to the corresponding Yukawa couplings
and hence $A_\nu$ is suppressed in our model. To zeroth order in
$\epsilon$ that implies first of all that CP even and CP odd sneutrinos have the
same mass and
\begin{equation} 
m^2_{\tilde{\nu}^R} \approx m^2_{\tilde{\nu}^I} \approx \left( 
\begin{array}{ccc}
\Re (M_{\tilde{L}}^2 ) + \tfrac{1}{2} M_Z^2 \cos (2 \beta) & 0 & 0 \\ 
0 & \Re(M_{\tilde{N}^c}^2  + \mu_{NS} \mu_{NS}^\dagger) & \Re(b_{NS})\\ 
0  & \Re(b_{NS}^T) & \Re(M_{\tilde{S}}^2  + \mu_{NS}^\dagger \mu_{NS}) \end{array} 
\right) \;.
\end{equation} 
At this point we do not know how $b_{NS}$ relates to $M_{\tilde{N}^c}^2  + \mu_{NS} \mu_{NS}^\dagger$
and $M_{\tilde{S}^c}^2  + \mu_{NS}^\dagger \mu_{NS}$ so that we do not know
if the mixing in this sector is large or small. Since we are working in a basis
where $\mu_{NS}$ is diagonal it is also reasonable to assume that $b_{NS}$
is diagonal. And hence the leading order contributions to cLFV are
expected to be induced by
$M_{\tilde{L}}^2$ and $M_{\tilde{N}^c}^2$ if the mixing between
right-handed sneutrinos and scalar singlets is small. If this mixing is large
$M_{\tilde{S}}^2$ could give sizeable cLFV in addition.
It is also interesting to note that the mixing between left-handed sneutrinos
and the new singlets is expected to be rather small due to a suppression
by smallish Yukawa couplings.

This concludes our discussion for charged-lepton flavor violation in our
model. We have seen that the non-SUSY contributions are much smaller
than current bounds and the SUSY contributions can in principle be
suppressed by pushing SUSY partners to the heavy limit. 
With the constraints given in this work so far, 
the SUSY partners do not necessarily have to be light.    
Nevertheless, this could change 
once we discuss potential dark matter candidates, for instance,
and some interesting non-trivial interplay might emerge.

\section{Summary and conclusions}
\label{sec:summary}

In this work, we have proposed a minimal supersymmetric inverse seesaw model 
with only two generations of right-handed neutrinos $\hat{N}^c$,  
two generations of singlet fields $\hat{S}$, one symmetry breaking singlet field
$\hat{X}$ and a $Z_6$ symmetry compared to the MSSM.
With the $Z_6$ charge assignments listed in Table~\ref{tab:Model}
we have successfully forbidden some unwanted terms 
(e.g.\ $\hat L \hat H_u \hat S$, $\hat H_u \hat H_d \hat N^c$ in the
superpotential) and retained those (e.g.\ $\hat L \hat H_u \hat N^c$)
relevant for generating the neutrino mass. In our model we also have an
intimate relation between the scale of electroweak symmetry breaking (or SUSY breaking)
and the mass scales in the neutrino sector avoiding a common
ad-hoc assumption in many models. This makes our model very well motivated
and attractive from a model building point of view.

We have studied three different types of our model according to the
mass hierarchy among $M_S$, $M_D$ and $\mu_{NS}$.
In all three types, we find three light active neutrino states with one neutrino
being massless due to our minimality assumption. The mixing angles are
consistent with current oscillation data which can be easily understood
from the reformulation of our leading order light neutrino mass matrix
in terms of the Casas-Ibarra parametrization as we have discussed.
So we fulfill the minimal requirement of any neutrino mass model.

Due to the fact that the neutrino mixing matrix is now enlarged the
3$\times$3 matrix tested in oscillations is expected to be non-unitary but our
estimates for this effect is far below current bounds.
Furthermore, since in our model the light active neutrinos are Majorana particles
we predict neutrinoless double beta decay with an effective mass
$\mathcal{O}(1)$~meV and $\mathcal{O}(10)$~meV for the normal
and inverted hierarchy neutrino masses, respectively. These tiny
numbers are experimentally challenging but on the other hand
a confirmed positive signal for non-unitarity in the mixing matrix
or neutrinoless double beta decay in the near future would immediately
challenge our model in its minimal version.

Furthermore, we have shown qualitatively 
that charged lepton flavor violation with both SUSY and non-SUSY
contributions can easily be below the current experimental bounds.
The non-SUSY contributions are in fact far below current and future
bounds and the SUSY contributions are under control since up to this
point the SUSY breaking parameters can easily be in the few
to several TeV region suppressing cLFV sufficiently.

This might nevertheless change if we include further constraints.
Our model has a rich dark matter and collider phenomenology
which is beyond the scope of the current work but will be discussed in
future publications in detail. Still we would like to use this
opportunity to make a few general comments.

\begin{enumerate}
\item 
Although we did not impose $R$-parity in our model, the 
conventional $R$-parity violating operators, such as $LQD^c,\,LLE^c, U^c D^c D^c$, and
$L H_u$ are not allowed by the $Z_6$ symmetry.

\item
 We have shown that the sneutrinos from the superfields $\hat L$,
$\hat N^c$, and $\hat S$ can all be mixed. In such a setup the lightest sneutrino could
be a dark matter candidate. 
In the conventional MSSM, if the LSP is a left-handed sneutrino,
it has been ruled out already by current direct detection 
experiments because of its large elastic cross-section with 
nuclei via $Z$-boson exchange.
However, in our current model the left-handed sneutrinos can mix with
the right-handed sneutrinos and extra singlets. In such a case, the elastic
scattering cross section can be suppressed or diluted to satisfy 
direct detection constraints. 

\item
The additional fermionic states which we have introduced are all expected
to have masses around a TeV. This is around the corner from the
collider physics point of view and the model can be tested in current
and upcoming experiments. This is indeed the main motivation for many
low scale seesaw models while here it is just another appealing feature.

\item
The presence of a number of sneutrinos coming from the mixing
of $\tilde{\nu}_L$, $\tilde{N}^c$, and $\tilde{S}$ would distinguish
the current model from the conventional MSSM. The sneutrinos can be 
directly produced via $Z$-boson exchange, or indirectly in some 
subsequent decays of heavier SUSY particles. If the mixing angle
among the inert sneutrinos and the left-handed sneutrino is sufficiently 
small, the decay of the heavier sneutrinos may be prolonged such that
it travels a distance without any tracks but suddenly decays 
with a vertex at some distance from the primary interaction point.
Such an event may be detectable using the MATHUSLA detector \cite{mathusla}.

\item
Any attempt towards a complete model of particle physics should also
provide a dynamical mechanism for baryogenesis. The seesaw mechanism
offers with Leptogenesis \cite{Fukugita:1986hr} an extremely popular
solution for this. If this baryogenesis mechanism or another mechanism
works in our model is left for another future study.

\end{enumerate}

In summary our model provides a novel and rather minimal approach to supersymmetric
inverse seesaw models which comes in three variants with distinct phenomenologies
already in the lepton sector alone. Similar to any low scale seesaw model and in particular
supersymmetric models our model provides an incredibly rich phenomenology from
which we have just touched the tip of the iceberg. In fact, it can be tested at the
energy, the intensity and the precision frontier as we have started to discuss
here but will be discussed in greater detail in future work.

\section*{Acknowledgements}

This research was supported in parts by the Ministry of Science and Technology (MoST) of Taiwan
under Grant No.\ MOST-105-2112-M-007-028-MY3.
J.C.\ was supported by the National Research Foundation of Korea (NRF) grant No.\ NRF-2016R1E1A1A01943297.

\appendix

\section{Explicit expressions for mixing of light and heavy neutrinos}
\label{sec:explicit}

Since the explicit expressions for the mixing between
the light and the heavy neutrinos, $W$, are rather
long and not insightful we present them here in the
appendix. Our expressions are unitary up to order
$\epsilon^2$ which is sufficient for our purposes.
We also use $\eta$ to label the order
in $\epsilon$ explicitly throughout the appendix.

To make the expressions shorter we define the abbreviations
\begin{equation}
 A=M_D\mu_{NS}^{-1},\quad B=M_S\mu_{NS}^{-1}, \quad D=M^{T}_{D}M^{\ast}_{D},\quad E=(\mu_{NS}^*)^{-1}\mu_{NS}^{-1} \; .
\end{equation}
For the ISS type I the mixing matrix elements are
\begin{align}
 (W_{\text{I}})_{11} &=1 - 1/2 \, \eta_{\text{I}}^2 A \, A^\dagger + \eta_{\text{I}}^4 (A \, A^\dagger)^2 
  - 1/4 \, \eta_{\text{I}}^6 (A A^\dagger)^3 + 1/2 \, \eta_{\text{I}}^8 A \, B \, B^\dagger \, A^\dagger \, A \, A^\dagger \nonumber\\
 &+ \eta_{\text{I}}^8 A \, B \, E \, D \, B^\dagger \, A^\dagger + 1/4 \, \eta_{\text{I}}^8 (A \, A^\dagger)^4
   - 1/4 \, \eta_{\text{I}}^{10} A \, A^\dagger \, A \, B \, B^\dagger \, A^\dagger \, A \, A^\dagger \nonumber\\
 &- 1/2 \, \eta_{\text{I}}^{10} A \, A^\dagger \, A \, B \, E \, D \, B^\dagger \, A^\dagger
 - 1/8 \, \eta_{\text{I}}^{10} (A \, A^\dagger)^5  \;,\\
 (W_{\text{I}})_{22} &= 1 + 1/2 \, \eta_{\text{I}}^8 \, B^\dagger \, A^\dagger \, A \, A^\dagger \, A \, B + 
 \eta_{\text{I}}^8 \, B^\dagger \, A^\dagger \, A \, B \, D \, E \;, \\
 (W_{\text{I}})_{33} &= 1 - 1/2 \, \eta_{\text{I}}^2 \, A^\dagger \, A
 + \eta_{\text{I}}^4 (A^\dagger \, A)^2 
 - 1/4 \, \eta_{\text{I}}^6 (A^\dagger \, A)^3  + 1/4 \, \eta_{\text{I}}^8 (A^\dagger \, A)^4
 - 1/8 \, \eta_{\text{I}}^{10} (A^\dagger \, A)^5 \;, \\
 (W_{\text{I}})_{12} &=  \eta_{\text{I}}^3 \, A \, B - \eta_{\text{I}}^5 \, A \, A^\dagger \, A \, B - \eta_{\text{I}}^5 \,A \, B \, D \, E    
    + 1/4 \, \eta_{\text{I}}^7 \, (A \, A^\dagger)^2  A \, B + 1/2 \eta_{\text{I}}^7 \, A \, A^\dagger \, A \, B \, D \, E \nonumber\\
    &- 1/2 \, \eta_{\text{I}}^9 \, (A \, A^\dagger)^3 A \, B
    - \eta_{\text{I}}^9 \, (A \, A^\dagger)^2 A \, B \, D \, E \;,\\
 (W_{\text{I}})_{13} &= -\eta_{\text{I}} \, A + \eta_{\text{I}}^3 \, A \, A^\dagger \, A 
  - 3/4 \, \eta_{\text{I}}^5 \, (A \, A^\dagger)^2 A + 1/8 \, \eta_{\text{I}}^7 \, (A \, A^\dagger)^3 A
  - 1/4 \, \eta_{\text{I}}^9 \, (A \, A^\dagger)^4 A \;, \\
 (W_{\text{I}})_{23} &= 1/4 \, \eta_{\text{I}}^8 \, B^\dagger \, (A^\dagger \, A)^3 \;,\\
 (W_{\text{I}})_{21} &= - \eta_{\text{I}}^3 \, B^\dagger \, A^\dagger + 1/2 \, \eta_{\text{I}}^5 \, B^\dagger \, A^\dagger \, A \, A^\dagger + \eta_{\text{I}}^5 \, E \, D \, B^\dagger \, A^\dagger\;, \\
 (W_{\text{I}})_{31} &= \eta_{\text{I}} \, A^\dagger - \eta_{\text{I}}^3 \, A^\dagger \, A \, A^\dagger
   + 3/4 \, \eta_{\text{I}}^5 \, (A^\dagger \, A)^2 A^\dagger - 1/8 \, \eta_{\text{I}}^7 \, (A^\dagger \, A)^3 A^\dagger \nonumber\\
  &+ 1/2 \, \eta_{\text{I}}^9 \, A^\dagger \, A \, B \, B^\dagger \, A^\dagger \, A \, A^\dagger + \eta_{\text{I}}^9 \, A^\dagger \, A \, B \, E \, D \, B^\dagger \, A^\dagger
   + 1/4 \, \eta_{\text{I}}^9 \, (A^\dagger \, A)^4 A^\dagger \;,\\
 (W_{\text{I}})_{32} &= \eta_{\text{I}}^4 \, A^\dagger \, A \, B - 1/2 \, \eta_{\text{I}}^6 \, (A^\dagger \, A)^2 \, B
   - \eta_{\text{I}}^6 \, A^\dagger \, A \, B \, D \, E
   + 1/2 \, \eta_{\text{I}}^8 \, (A^\dagger \, A)^3  B \nonumber\\
  &+ \eta_{\text{I}}^8 \, (A^\dagger \, A)^2 B \, D \, E
   - 1/4 \, \eta_{\text{I}}^{10} \, (A^\dagger \, A)^4 B
   - 1/2 \, \eta_{\text{I}}^{10} \, (A^\dagger \, A)^3 B \, D \, E \;,
\end{align}
where we have quoted for convenience the orders in $\eta_{\text{I}}$ explicitly.

For the ISS type II the mixing matrix elements are
\begin{align}
 (W_{\text{II}})_{11} &= 1 - 1/2 \, \eta_{\text{II}}^2 \, A \, A^\dagger + \eta_{\text{II}}^4 \, (A \, A^\dagger)^2
   + 1/2 \, \eta_{\text{II}}^6 \, A \, B \, B^\dagger \, A^\dagger \, A \, A^\dagger 
   + \eta_{\text{II}}^6 \, A \, B \, E \, D \, B^\dagger \, A^\dagger \nonumber\\
   &- 1/4 \, \eta_{\text{II}}^8 \, A \, A^\dagger \, A \, B \, B^\dagger \, A^\dagger \, A \, A^\dagger - 1/2 \, \eta_{\text{II}}^8 \, A \, A^\dagger \, A \, B \, E \, D \, B^\dagger \, A^\dagger \;, \\
 (W_{\text{II}})_{22} &= 1 + 1/2 \, \eta_{\text{II}}^6 B^\dagger \, A^\dagger \, A \, A^\dagger \, A \, B
   + \eta_{\text{II}}^6 \, B^\dagger \, A^\dagger \, A \, B \, D \, E \;, \\
 (W_{\text{II}})_{33} &=1 - 1/2 \, \eta_{\text{II}}^2 \, A^\dagger \, A + \eta_{\text{II}}^4 A^\dagger \, A \, A^\dagger \, A \;, \\
 (W_{\text{II}})_{12} &= \eta_{\text{II}}^2 \, A \, B - \eta_{\text{II}}^4 A \, A^\dagger \, A \, B 
   - \eta_{\text{II}}^4 \, A \, B \, D \, E
   + 1/4 \, \eta_{\text{II}}^6 \, (A \, A^\dagger)^2 A \, B \nonumber\\
  &+ 1/2 \, \eta_{\text{II}}^6 \, A \, A^\dagger \, A \, B \, D \, E
   - 1/2 \, \eta_{\text{II}}^8 \, (A \, A^\dagger)^3 A \, B 
   - \eta_{\text{II}}^8 \, (A \, A^\dagger)^2  A \, B \, D \, E  \;, \\
 (W_{\text{II}})_{13} &= - \eta_{\text{II}}  \, A + \eta_{\text{II}}^3 \, A \, A^\dagger \, A  - 
 1/2 \, \eta_{\text{II}}^5 \, (A \, A^\dagger)^2 A  \;, \\
 (W_{\text{II}})_{23} &= - \eta_{\text{II}}^5 \, B^\dagger  (A^\dagger \, A)^2 \;, \\
 (W_{\text{II}})_{21} &=-\eta_{\text{II}}^2 \, B^\dagger \, A^\dagger + 1/2 \, \eta_{\text{II}}^4 \, B^\dagger \, A^\dagger \, A \, A^\dagger + \eta_{\text{II}}^4  \, E \, D \, B^\dagger \, A^\dagger \;, \\
 (W_{\text{II}})_{31} &= \eta_{\text{II}} \, A^\dagger - \eta_{\text{II}}^3 \, A^\dagger \, A \, A^\dagger 
    + 1/2 \, \eta_{\text{II}}^5 \, (A^\dagger \, A)^2 A^\dagger 
    + 1/2 \, \eta_{\text{II}}^7 \, A^\dagger \, A \, B \, B^\dagger \, A^\dagger \, A \, A^\dagger \nonumber\\
   &+ \eta_{\text{II}}^7 \, A^\dagger \, A \, B \, E \, D \, B^\dagger \, A^\dagger \;, \\
 (W_{\text{II}})_{32} &= \eta_{\text{II}}^3 \, A^\dagger \, A \, B - 1/2 \, \eta_{\text{II}}^5 (A^\dagger \, A)^2 B
   -  \eta_{\text{II}}^5 \, A^\dagger \, A \, B \, D \, E + 1/2 \, \eta_{\text{II}}^7 \, (A^\dagger \, A)^3  B \nonumber\\
  &+ \eta_{\text{II}}^7 \, (A^\dagger \, A)^2  B \, D \, E
  - 1/4 \, \eta_{\text{II}}^9 \, (A^\dagger \, A)^4 B - 1/2 \, \eta_{\text{II}}^9 \, (A^\dagger \, A)^3  B \, D \, E \;,       
\end{align}
where we have quoted for convenience the orders in $\eta_{\text{II}}$ explicitly.

For the ISS type III the mixing matrix elements are
\begin{align}
 (W_{\text{III}})_{11} &= 1 \;, \\
 (W_{\text{III}})_{22} &= 1 \;, \\
 (W_{\text{III}})_{33} &= 1 \;, \\
 (W_{\text{III}})_{12} &= \eta_{\text{III}}^3\, A\, B  \;, \\
 (W_{\text{III}})_{13} &= -\eta_{\text{III}}^2\, A  \;, \\
 (W_{\text{III}})_{23} &= \mathcal{O}(\eta_{\text{III}}^{13} ) \;, \\
 (W_{\text{III}})_{21} &= - \eta_{\text{III}}^3\, B^\dagger \, A^\dagger \;, \\
 (W_{\text{III}})_{31} &= -\eta_{\text{III}}^2\, A^\dagger  \;, \\
 (W_{\text{III}})_{32} &= \eta_{\text{III}}^5\, A^{\dagger}\, A\, B \;,       
\end{align}
where we have quoted for convenience the orders in $\eta_{\text{III}}$ explicitly.


\begin{thebibliography}{99}
 
 
  \bibitem{mssm}
H.~E.~Haber and G.~L.~Kane,
  Phys.\ Rept.\  {\bf 117}, 75 (1985).

\bibitem{osc}
See for example, 
D.~V.~Forero, M.~Tortola and J.~W.~F.~Valle,
  Phys.\ Rev.\ D {\bf 90}, no. 9, 093006 (2014)
  [arXiv:1405.7540 [hep-ph]].

\bibitem{rp}
M.~Hirsch, M.~A.~Diaz, W.~Porod, J.~C.~Romao and J.~W.~F.~Valle,
  Phys.\ Rev.\ D {\bf 62}, 113008 (2000)
  Erratum: [Phys.\ Rev.\ D {\bf 65}, 119901 (2002)]
  [hep-ph/0004115].

\bibitem{seesaw}
  P.~Minkowski,
  Phys.\ Lett.\ B {\bf 67}, 421 (1977); 
T. Yanagida, in Proceedings of the Workshop on Unified Theory and Baryon Number of the Universe, edited by O. Sawada and A. Sugamoto (KEK, Tsukuba, Ibaraki 305- 0801 Japan, 1979) p. 95; 
  T.~Yanagida,
  Prog.\ Theor.\ Phys.\  {\bf 64}, 1103 (1980); 
  M. Gell-Mann, P. Ramond, and R. Slansky, in Supergravity, edited by P. van Niewwenhuizen and D. Freedman (North Holland, Amsterdam, 1979)
  [arXiv:1306.4669 [hep-th]]; 
P.~Ramond, 
in {\em Talk given at the Sanibel Symposium}, 
Palm Coast, Fla., Feb.~25-Mar.~2, 1979, preprint CALT-68-709
(retroprinted as hep-ph/9809459); 
S.~L.~Glashow,
in {\em Proc. of the Carg\'ese  Summer Institute on Quarks and Leptons},
Carg\'ese, July 9-29, 1979, 
eds. M.~L\'evy et. al, , (Plenum, 1980, New York), p707;
R.~N.~Mohapatra and G.~Senjanovic,
Phys.\ Rev.\ Lett.\  {\bf 44} (1980) 912;
J.~Schechter and J.~W.~Valle,
Phys.\ Rev.\ D {\bf 25} (1982) 774.


\bibitem{inverse}
  R.~N.~Mohapatra,
  Phys.\ Rev.\ Lett.\  {\bf 56} (1986) 561;
  R.~N.~Mohapatra and J.~W.~F.~Valle,
  Phys.\ Rev.\ D {\bf 34} (1986) 1642.

\bibitem{Das:2014jxa}
  A.~Das, P.~S.~Bhupal Dev and N.~Okada,
  Phys.\ Lett.\ B {\bf 735} (2014) 364
  [arXiv:1405.0177 [hep-ph]].
  
\bibitem{Das:2015toa} 
  A.~Das and N.~Okada,
  Phys.\ Rev.\ D {\bf 93}, no. 3, 033003 (2016)
  [arXiv:1510.04790 [hep-ph]].  

  
\bibitem{100tev}
M.~L.~Mangano {\it et al.},
  arXiv:1607.01831 [hep-ph].
  
  
\bibitem{Das:2012ze}
  A.~Das and N.~Okada,
  Phys.\ Rev.\ D {\bf 88} (2013) 113001
  [arXiv:1207.3734 [hep-ph]].

\bibitem{Antusch:2015mia}
  S.~Antusch and O.~Fischer,
  JHEP {\bf 1505} (2015) 053
  [arXiv:1502.05915 [hep-ph]].


\bibitem{Blondel:2014bra}
  A.~Blondel {\it et al.} [FCC-ee study Team],
  Nucl.\ Part.\ Phys.\ Proc.\  {\bf 273-275} (2016) 1883
  [arXiv:1411.5230 [hep-ex]].

\bibitem{Antusch:2016ejd}
  S.~Antusch, E.~Cazzato and O.~Fischer,
  Int.\ J.\ Mod.\ Phys.\ A {\bf 32} (2017) no.14,  1750078
  [arXiv:1612.02728 [hep-ph]].


%
\bibitem{Atre:2009rg}
  See for example,
  A.~Atre, T.~Han, S.~Pascoli and B.~Zhang,
  JHEP {\bf 0905} (2009) 030
  [arXiv:0901.3589 [hep-ph]].

\bibitem{Arganda:2015naa}
  E.~Arganda, M.~J.~Herrero, X.~Marcano and C.~Weiland,
  Phys.\ Rev.\ D {\bf 93} (2016) no.5,  055010
  [arXiv:1508.04623 [hep-ph]].
  

\bibitem{Deppisch:2004fa} 
  F.~Deppisch and J.~W.~F.~Valle,
  Phys.\ Rev.\ D {\bf 72}, 036001 (2005)
  [hep-ph/0406040].



\bibitem{Arina:2008bb} 
  C.~Arina, F.~Bazzocchi, N.~Fornengo, J.~C.~Romao and J.~W.~F.~Valle,
  Phys.\ Rev.\ Lett.\  {\bf 101}, 161802 (2008)
  [arXiv:0806.3225 [hep-ph]].

\bibitem{BhupalDev:2012ru}
  P.~S.~Bhupal Dev, S.~Mondal, B.~Mukhopadhyaya and S.~Roy,
  JHEP {\bf 1209} (2012) 110
  [arXiv:1207.6542 [hep-ph]].

\bibitem{Hirsch:2009ra} 
  M.~Hirsch, T.~Kernreiter, J.~C.~Romao and A.~Villanova del Moral,
  JHEP {\bf 1001}, 103 (2010)
  [arXiv:0910.2435 [hep-ph]].

  
  

\bibitem{Khalil:2011tb} 
  S.~Khalil, H.~Okada and T.~Toma,
  JHEP {\bf 1107}, 026 (2011)
  [arXiv:1102.4249 [hep-ph]].
  


\bibitem{Hirsch:2011hg}
  M.~Hirsch, M.~Malinsky, W.~Porod, L.~Reichert and F.~Staub,
  JHEP {\bf 1202} (2012) 084
  [arXiv:1110.3037 [hep-ph]];
  M.~Hirsch, W.~Porod, L.~Reichert and F.~Staub,
  Phys.\ Rev.\ D {\bf 86} (2012) 093018
  [arXiv:1206.3516 [hep-ph]];
  V.~De Romeri and M.~Hirsch,
  JHEP {\bf 1212}, 106 (2012)
  [arXiv:1209.3891 [hep-ph]].


\bibitem{Kang:2011wb} 
  Z.~Kang, J.~Li, T.~Li, T.~Liu and J.~M.~Yang,
  Eur.\ Phys.\ J.\ C {\bf 76}, no. 5, 270 (2016)
  [arXiv:1102.5644 [hep-ph]].
  S.~L.~Chen and Z.~Kang,
  Phys.\ Lett.\ B {\bf 761} (2016) 296
  [arXiv:1512.08780 [hep-ph]].


\bibitem{Bazzocchi:2009kc} 
  F.~Bazzocchi, D.~G.~Cerdeno, C.~Munoz and J.~W.~F.~Valle,
  Phys.\ Rev.\ D {\bf 81}, 051701 (2010)
  [arXiv:0907.1262 [hep-ph]].




\bibitem{Dev:2009aw} 
  P.~S.~B.~Dev and R.~N.~Mohapatra,
  Phys.\ Rev.\ D {\bf 81}, 013001 (2010)
  [arXiv:0910.3924 [hep-ph]].


\bibitem{An:2011uq} 
  H.~An, P.~S.~B.~Dev, Y.~Cai and R.~N.~Mohapatra,
  Phys.\ Rev.\ Lett.\  {\bf 108}, 081806 (2012)
  [arXiv:1110.1366 [hep-ph]].

\bibitem{Cheung:2001sb}
  K.~m.~Cheung and O.~C.~W.~Kong,
  Phys.\ Rev.\ D {\bf 64} (2001) 095007
  [hep-ph/0101347].
 

\bibitem{Abada:2014vea}
  A.~Abada and M.~Lucente,
  Nucl.\ Phys.\ B {\bf 885} (2014) 651
  [arXiv:1401.1507 [hep-ph]].

\bibitem{Martin:1997ns}
  S.~P.~Martin,
  Adv.\ Ser.\ Direct.\ High Energy Phys.\  {\bf 21} (2010) 1
   [Adv.\ Ser.\ Direct.\ High Energy Phys.\  {\bf 18} (1998) 1]
  [hep-ph/9709356].

\bibitem{Mohapatra:1986bd}
  R.~N.~Mohapatra and J.~W.~F.~Valle,
  Phys.\ Rev.\ D {\bf 34} (1986) 1642.

  %
\bibitem{tHooft:1979rat}
  G.~'t Hooft,
  NATO Sci.\ Ser.\ B {\bf 59} (1980) 135.
  
%
\bibitem{Fernandez-Martinez:2016lgt}
  E.~Fernandez-Martinez, J.~Hernandez-Garcia and J.~Lopez-Pavon,
  JHEP {\bf 1608} (2016) 033
  [arXiv:1605.08774 [hep-ph]].  
  
\bibitem{Casas:2001sr}
  J.~A.~Casas and A.~Ibarra,
  Nucl.\ Phys.\ B {\bf 618} (2001) 171
  [hep-ph/0103065].   
  
%
\bibitem{Pas:2015eia}
H.~P\"{a}s and W.~Rodejohann,
  New J.\ Phys.\  {\bf 17} (2015) no.11,  115010
  [arXiv:1507.00170 [hep-ph]].


%
\bibitem{KamLAND-Zen:2016pfg}
  A.~Gando {\it et al.} [KamLAND-Zen Collaboration],
  Phys.\ Rev.\ Lett.\  {\bf 117} (2016) no.8,  082503
   Addendum: [Phys.\ Rev.\ Lett.\  {\bf 117} (2016) no.10,  109903]
  [arXiv:1605.02889 [hep-ex]].
  
%
\bibitem{Gonzalez-Garcia:2014bfa}
  M.~C.~Gonzalez-Garcia, M.~Maltoni and T.~Schwetz,
  JHEP {\bf 1411} (2014) 052
  [arXiv:1409.5439 [hep-ph]].  
  
%
\bibitem{Mohapatra:1986su}
  R.~N.~Mohapatra,
  Phys.\ Rev.\ D {\bf 34} (1986) 3457.
  
%
\bibitem{Faessler:2014kka}
  A.~Faessler, M.~Gonz\'alez, S.~Kovalenko and F.~\v{S}imkovic,
  Phys.\ Rev.\ D {\bf 90} (2014) no.9,  096010
  [arXiv:1408.6077 [hep-ph]].


%
\bibitem{Haba:2016lxc}
  N.~Haba, H.~Ishida and Y.~Yamaguchi,
  JHEP {\bf 1611} (2016) 003
  [arXiv:1608.07447 [hep-ph]].  
  
\bibitem{Cheng:1980tp}
  T.~P.~Cheng and L.~F.~Li,
  Phys.\ Rev.\ Lett.\  {\bf 45} (1980) 1908.
  
\bibitem{Petcov:1976ff}
  S.~T.~Petcov,
  Sov.\ J.\ Nucl.\ Phys.\  {\bf 25} (1977) 340
   [Yad.\ Fiz.\  {\bf 25} (1977) 641]
   Erratum: [Sov.\ J.\ Nucl.\ Phys.\  {\bf 25} (1977) 698]
   Erratum: [Yad.\ Fiz.\  {\bf 25} (1977) 1336];
  
\bibitem{TheMEG:2016wtm}
  A.~M.~Baldini {\it et al.} [MEG Collaboration],
  Eur.\ Phys.\ J.\ C {\bf 76} (2016) no.8,  434
  [arXiv:1605.05081 [hep-ex]].
  
\bibitem{Hisano:1995cp}
  J.~Hisano, T.~Moroi, K.~Tobe and M.~Yamaguchi,
  Phys.\ Rev.\ D {\bf 53} (1996) 2442
  [hep-ph/9510309];
  J.~Hisano, T.~Moroi, K.~Tobe, M.~Yamaguchi and T.~Yanagida,
  Phys.\ Lett.\ B {\bf 357} (1995) 579
  [hep-ph/9501407].
  
\bibitem{Staub:2008uz}
  F.~Staub,
  arXiv:0806.0538 [hep-ph];
  F.~Staub,
  Comput.\ Phys.\ Commun.\  {\bf 185} (2014) 1773
  [arXiv:1309.7223 [hep-ph]].
  
  
\bibitem{Abada:2014kba}
  A.~Abada, M.~E.~Krauss, W.~Porod, F.~Staub, A.~Vicente and C.~Weiland,
  JHEP {\bf 1411} (2014) 048
  [arXiv:1408.0138 [hep-ph]].
  

\bibitem{mathusla}
  J.~P.~Chou, D.~Curtin and H.~J.~Lubatti,
  Phys.\ Lett.\ B {\bf 767} (2017) 29
  [arXiv:1606.06298 [hep-ph]];
  D.~Curtin and M.~E.~Peskin,
  arXiv:1705.06327 [hep-ph].

\bibitem{Fukugita:1986hr}
  M.~Fukugita and T.~Yanagida,
  Phys.\ Lett.\ B {\bf 174} (1986) 45.
  

\end{thebibliography}
\end{document}